\documentclass[12pt, preprint, longabstract,usegraphicx]{article}
\usepackage{lscape}
\usepackage{ccaption}
\usepackage{comment}
\usepackage{graphicx}
\usepackage{deluxetable}
\usepackage[super,numbers,comma,sort]{natbib}

\newcommand{\sun}{\odot}
\newcommand{\nature}{{\it Nature}}
\newcommand{\science}{{\it Science}}
\newcommand{\apj}{{\it Astrophys. J.}}

\newcommand{\aj}{{\it Astron. J.}}

\newcommand{\aap}{{\it Astron. Astrophys.}}
\newcommand{\aaps}{{\it Astron. Astrophys. Suppl.}}
\newcommand{\mnras}{{\it Mon. Not. R. Astron. Soc.}}
\newcommand{\pasp}{{\it  Pub. Astron. Soc. Pacific}}
\newcommand\msun {M_\odot}

\newcommand\gtsima{$\; \buildrel >\over\sim \;$}
\newcommand\simgt{\lower.5ex\hbox{\gtsima}}
\newcommand\ltsima{$\; \buildrel <\over\sim \;$}
\newcommand\simlt{\lower.5ex\hbox{\ltsima}}

\begin{document}
%\title{Unbound or Distant Planetary Mass Population Detected by Gravitational Microlensing}
%\maketitle
%\author{}
\section*{\center Unbound or Distant Planetary Mass Population Detected by Gravitational Microlensing}
T. Sumi$^{1,2,3}$, K. Kamiya$^{1,3}$, A. Udalski$^{4,5}$, D.P. Bennett$^{1,6}$, I.A. Bond$^{1,7}$,
F. Abe$^{1,3}$, C.S. Botzler$^{1,8}$, A. Fukui$^{1,3}$, K. Furusawa$^{1,3}$, J.B. Hearnshaw$^{1,9}$, 
Y. Itow$^{1,3}$, P.M. Kilmartin$^{1,10}$, A. Korpela$^{1,11}$, W. Lin$^{1,7}$, C.H. Ling$^{1,7}$, 
K. Masuda$^{1,3}$, Y. Matsubara$^{1,3}$, N. Miyake$^{1,3}$, M. Motomura$^{1,3}$, Y. Muraki$^{1,12}$, 
M. Nagaya$^{1,3}$, S. Nakamura$^{1,3}$, K. Ohnishi$^{1,13}$, T. Okumura$^{1,3}$, Y.C. Perrott$^{1,14}$, 
N. Rattenbury$^{1,8}$, To. Saito$^{1,15}$, T. Sako$^{1,3}$, D.J. Sullivan$^{1,11}$, W.L. Sweatman$^{1,7}$,
P.J. Tristram$^{1,8}$, P.C.M. Yock$^{1,8}$, M.K. Szyma\'nski$^{4,5}$, M. Kubiak$^{4,5}$, 
G. Pietrzy\'nski$^{4,5,16}$, R. Poleski$^{4,5}$, I. Soszy\'nski$^{4,5}$, \L. Wyrzykowski$^{4,17}$, 
K. Ulaczyk$^{4,5}$ \\\\
{\it
$^{1}$ Microlensing Observations in Astrophysics (MOA) Collaboration.\\
$^{2}$ Department of Earth and Space Science, Osaka University, Osaka 560-0043, Japan.\\
$^{3}$ Solar-Terrestrial Environment Laboratory, Nagoya University, Nagoya 464-8601, Japan.\\
$^{4}$ Optical Gravitational Lens Experiment (OGLE) Collaboration.\\
$^{5}$ Warsaw University Observatory, Al. Ujazdowskie 4, 00-478 Warszawa, Poland.\\
$^{6}$ Department of Physics, University of Notre Dame, Notre Dame, IN 46556, USA.\\
$^{7}$ Institute of Information and Mathematical Sciences, Massey University,
Private Bag 102-904, North Shore Mail Centre, Auckland 0745, New Zealand.\\
$^{8}$ Department of Physics, University of Auckland, Private Bag 92019, Auckland 1142, New Zealand.\\
$^{9}$ Dept. of Physics and Astronomy, University of Canterbury, Christchurch 8140, New¿Zealand.\\
$^{10}$ Mt. John University Observatory, University of Canterbury, P.O. Box 56, Lake Tekapo 8770, New Zealand.\\
$^{11}$ School of Chemical and Physical Sciences, Victoria University, Wellington 6140, New Zealand.\\
$^{12}$ Department of Physics, Konan University, Nishiokamoto 8-9-1, Kobe 658-8501, Japan.\\
$^{13}$ Nagano National College of Technology, Nagano 381-8550, Japan.\\
$^{14}$ Cavendish Laboratory, Cambridge University, J. J. Thomson Avenue, CB3 0HE Cambridge, UK. \\
$^{15}$ Tokyo Metropolitan College of Industrial Technology, Tokyo 116-8523, Japan.\\
$^{16}$ Universidad de Concepci\'on, Departamento de Astronomia, Casilla 160--C, Concepci\'on, Chile.\\
$^{17}$ Institute of Astronomy, Cambridge University, Madingley Rd., CB3 0HA Cambridge, UK.\\
}

{\bf
Since 1995, more than 500 exoplanets have been detected using different 
techniques\citep{may95,Schneider2011}, of which 11 were detected with 
gravitational microlensing\citep{Gaudi2010,Sumi2010}. Most of these are 
gravitationally bound to their host stars. There is some evidence of 
free-floating planetary mass objects in young star-forming 
regions\cite{ZapateroOsorio2000,Burgess2009,Quanz2010,Marsh2010}, 
but these objects are limited to massive objects of 3 to 15 Jupiter masses 
with large uncertainties in photometric mass estimates and their abundance.  
Here, we report the discovery of a population of unbound or distant Jupiter-mass 
objects, which are almost twice ($1.8_{-0.8}^{+1.7}$) as common as main-sequence 
stars, based on two years of gravitational microlensing survey observations 
toward the Galactic Bulge. These planetary-mass objects have no host stars 
that can be detected within about ten astronomical units by gravitational 
microlensing. However a comparison with constraints from direct 
imaging\citep{Lafreniere2007} suggests that most of these planetary-mass objects 
are not bound to any host star. An abrupt change in the mass function at about 
a Jupiter mass favours the idea that their formation process is different from 
that of stars and brown dwarfs. They may have formed in proto-planetary disks 
and subsequently scattered into unbound or very distant orbits. 
}\\

In a gravitational microlensing event, a foreground lens object is detected 
as a result of the characteristic magnification of a background source star 
as it passes behind the gravitational field of the lens\cite{pac86}. The lens 
object is detected by means of its mass and not its luminosity. The duration 
of the magnification is parameterized by the Einstein radius crossing time,
$t_{\rm E}\sim\sqrt{M/M_{\rm J}}$ days, where $M_{\rm J} = 9.5\times10^{-4} M_\sun$¿
 is Jupiter¿s mass. Thus, microlensing can detect faint planetary mass objects 
---which are either unbound to any host star\cite{Liebes1964,Bennett1997}
or are in very wide orbits\cite{DiStefano1999a} --- as short-timescale events 
with $t_{\rm E} < 2$ days. Although $t_{\rm E}$ also depends on the distance 
and transverse velocity of the lens (see Supplementary Information), the 
observed $t_{\rm E}$ distribution can be used as a statistical probe of the 
mass function of the lens objects because the spatial and velocity distributions 
in the Galactic disk and bulge are reasonably well known. 

The Microlensing Observations in Astrophysics (MOA)\cite{sumi03} and Optical 
Gravitational Lensing Experiment (OGLE)\cite{Udalski2003} groups both conduct 
microlensing surveys toward the Galactic bulge. The second phase of MOA, 
called MOA-II, carries out a high cadence photometric survey of 50 million 
stars in bulge fields with a cadence of 10-50 minutes. This strategy enables 
MOA to detect very short events with $t_{\rm E} < 2$ days, which were quite 
rare in previous microlensing surveys that had lower 
cadences\cite{Bennett1997,sumi03,uda94,alc00a}.

In this analysis of the 2006-2007 MOA-II data set, light curves of genuine 
microlensing events were distinguished from intrinsic variables and artefacts by 
several empirical criteria, which have been developed in previous microlensing 
surveys\cite{sumi03,Udalski2003,uda94,alc00a,Hamadache2006,Sumi2006}.  
The light curves must have a single brightening episode of more than three 
consecutive significant data points with a constant baseline, and should 
be well fitted by a theoretical microlensing model\cite{pac86} with a well 
constrained $t_{\rm E}$ (see the Supplementary Information). 

\begin{figure*}
%\epsscale{0.5}
\includegraphics[angle=-90,scale=0.33,keepaspectratio]{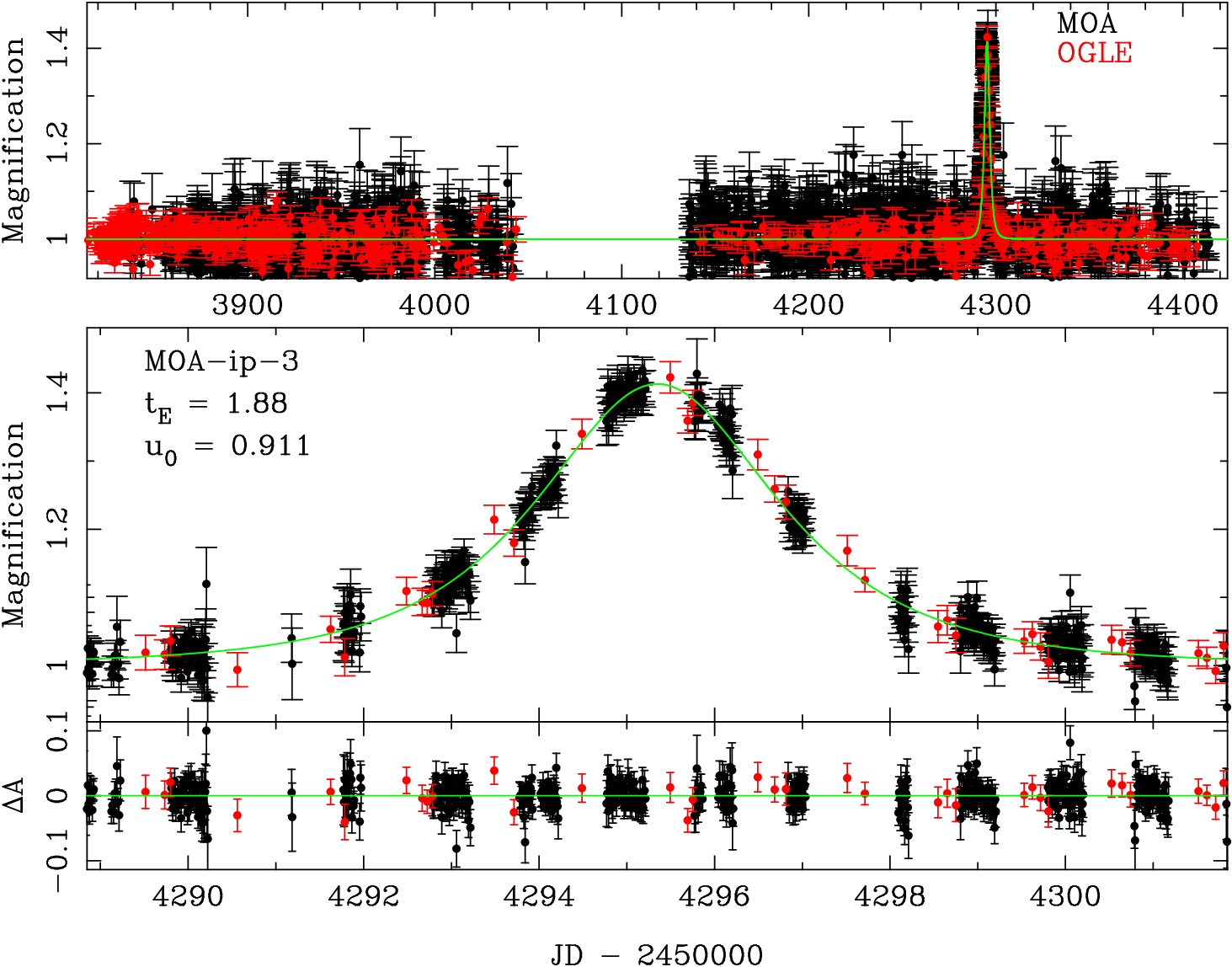}
\includegraphics[angle=-90,scale=0.33,keepaspectratio]{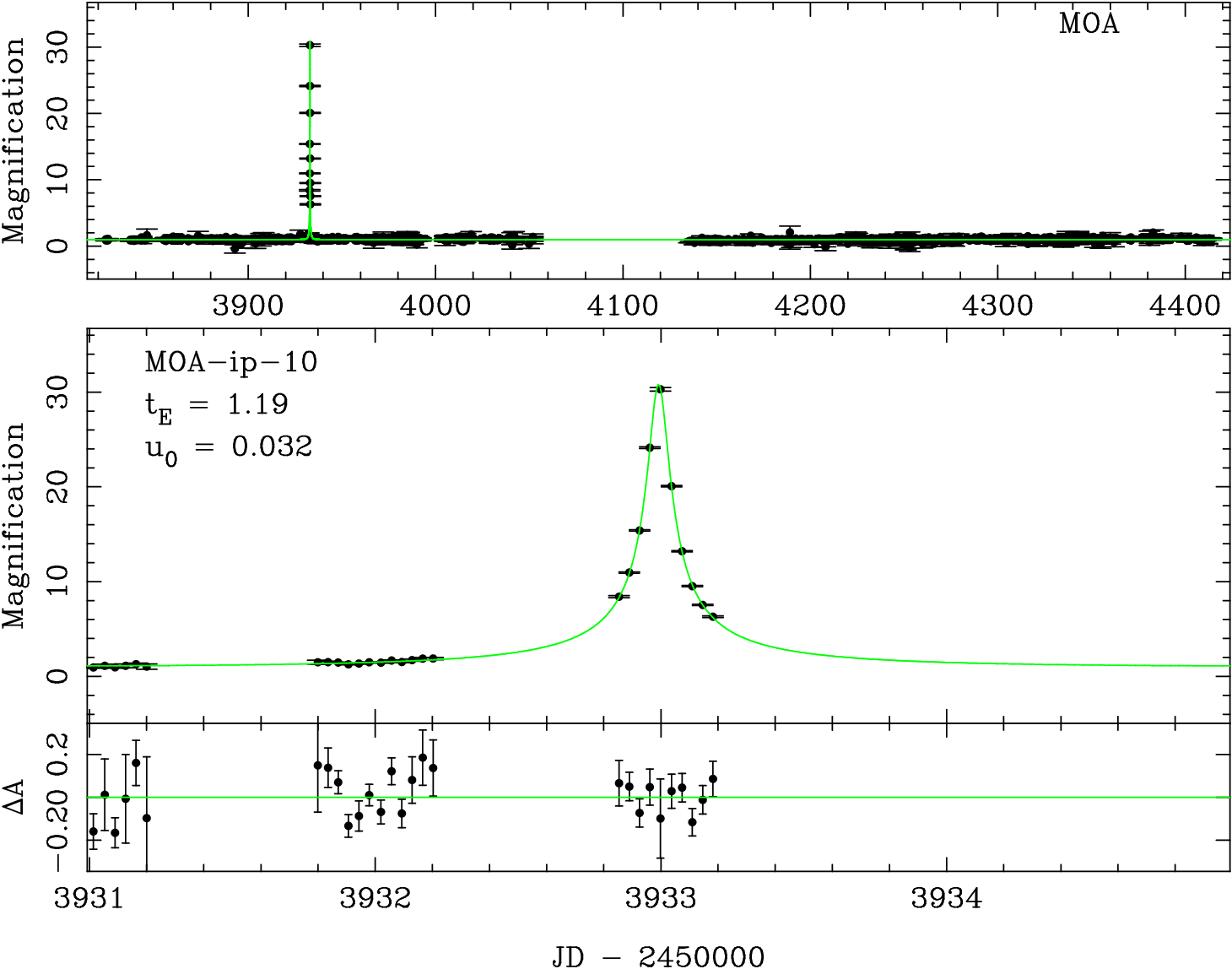}
\caption{
Light curves of event MOA-ip-3 and MOA-ip-10. These have the highest signal 
to noise ratio amongst the ten microlensing events with $t_{\rm E} < 2$ days 
(see Supplementary Fig. 1 for the others). MOA data are in black and OGLE 
data are in red with error bars indicating the s.e.m. The green lines 
represent the best-fit microlensing model light curves. For each event, 
the upper panel shows the full two-year light curve, the middle panel is 
a close-up of the light curve peak, and the bottom panel shows the residuals 
from the best-fit model in units of magnification,$\Delta A$. $u_0$ indicates 
the source-lens impact parameter in unit of the Einstein radius. The second 
phase of MOA, MOA-II, carried out a very high cadence photometric survey of 
50 million stars in 22 bulge fields (of 2.2 deg$^2$ each) with a 1.8m telescope 
at Mt. John Observatory in New Zealand. MOA detects 500-600 microlensing events 
with 8 months observation every year. In 2006-2007, MOA observed two central 
bulge fields every 10 minutes, and other bulge fields with a 50 minutes cadence, 
which result  $\sim$8250 and 1660-2980 images, respectively. This strategy 
enabled MOA to detect very short events with $t_{\rm E} < 2$ days. Since 2002, 
the OGLE-III survey has monitored the bulge with the 1.3-m Warsaw telescope 
at Las Campanas Observatory, Chile, with a smaller field-of-view but better 
astronomical seeing than MOA. The OGLE-III observing cadence was 1-2 observations 
per night, but the OGLE photometry is usually more precise and fills gaps 
in the MOA light curves due to the difference in longitude.
\label{fig:lc_main1}}
\end{figure*}

Although there are a thousand microlensing events in this sample, only 474 
well characterized events have passed our strict selection criteria. Ten of 
these events have $t_{\rm E} < 2$ days (see Fig. \ref{fig:lc_main1} and 
Table \ref{tbl:param}) thus indicating planetary-mass lenses. We have 
confirmed that this event sample has no significant contamination by possible 
background effects including: (1) cosmic-ray hits, (2) fast-moving objects, 
(3) cataclysmic variables (4) background supernovae, (5) binary microlensing 
events, and (6) microlensing by high-velocity stars and Galactic halo stellar 
remnants. For example, effect (1) is excluded because cosmic-rays never hit the 
same place in four consecutive images, microlensing model fits for effects (2) 
to (5) produce a high $\chi^2$ and unphysical values of parameters, and 
effect (6) is excluded by proper-motion and radial-velocity observations 
(see Supplementary Information). After the MOA event selection was complete, 
the MOA group requested additional independent light-curve data of these short 
events from the OGLE group. Seven of the ten events with $t_{\rm E} < 2$ days 
were also observed by OGLE-III\cite{Udalski2003}, and none of them have any 
other brightening in the eight-year OGLE-III light curves. For six of these 
seven events, there are OGLE data obtained during the lensing event that 
confirm the predictions of the MOA microlensing models. 
Thus, the OGLE data confirms the microlensing interpretation of these short events.

The detection efficiencies for this analysis were estimated with a Monte 
Carlo simulation\cite{sumi03,alc00a}. We simulated 20 million artificial 
events to evaluate the detection efficiency as a function of $t_{\rm E}$, 
yielding $\epsilon (t_{\rm E}) \sim$ 1\%, 3\%, 5\%, 10\%, 15\% and 10\% at 
$t_{\rm E} =$ 0.3, 1, 2, 10, 30 and 100 days, respectively. The details of 
the efficiency calculations and consistency tests of the selected event 
distribution are discussed in the Supplementary Information. 

\begin{figure}
\includegraphics[angle=-90,scale=0.6,keepaspectratio]{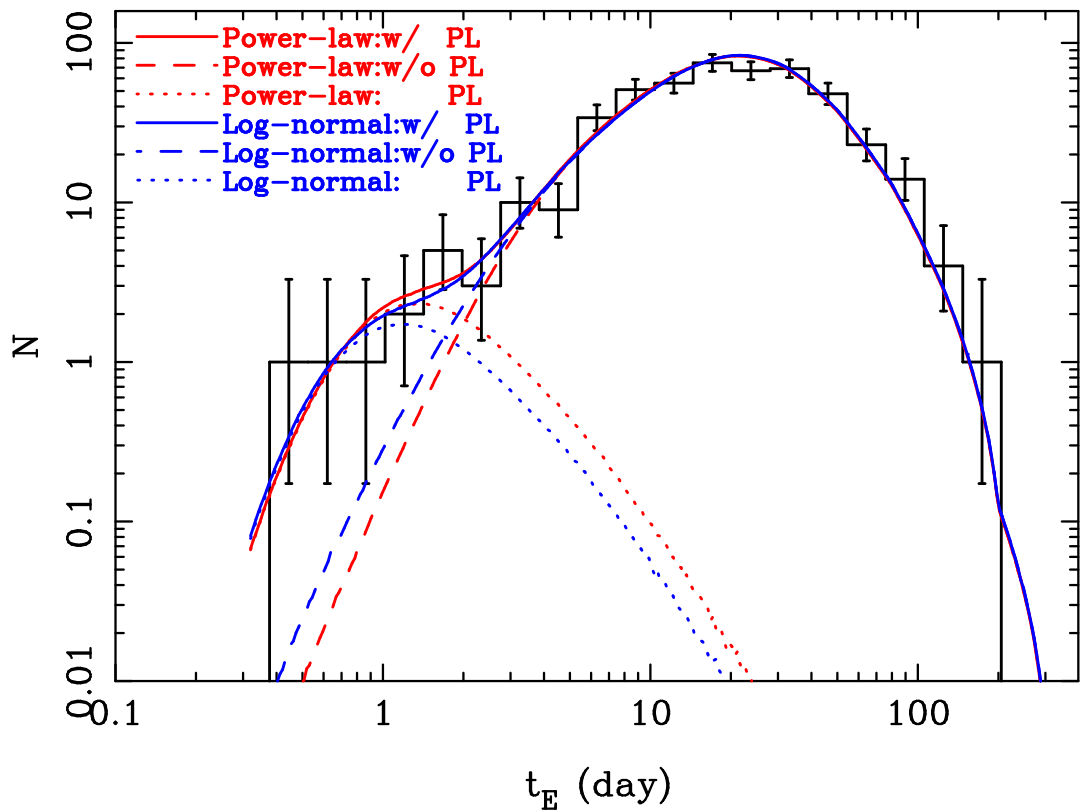}
\caption{
Observed and theoretical distributions of the event timescale, $t_{\rm E}$. 
The black histogram represents the number $N$ of observed 474 microlensing 
events in each bin with error bars indicating the s.e.m. The red and blue 
lines indicate the best-fit models with (1) the power-law and (2) log-normal
mass functions, respectively. For both mass functions, we assume that stars 
that were initially above 1 $M_\sun$ have evolved into stellar remnants --- 
white dwarfs, neutron stars or black holes depending on the initial mass. 
The number of remnants is determined by extending the upper main sequence 
power-law $\alpha_1 = 2.0$ to 100 $M_\sun$, and the final remnant mass 
distributions are given by Gaussians\cite{Gould2003} (see Supplementary Table 3). 
Each model is multiplied by the detection efficiencies. In each model, 
dashed lines indicate models for stellar, stellar remnant and brown dwarf 
populations, and the dotted lines represent the planetary-mass population (PL). 
Solid lines are the sums of these populations, and both models fit the data well.
\label{fig:tE}}
\end{figure}

The observed $t_{\rm E}$ distribution is compared to two mass function 
models in Fig. \ref{fig:tE}. A model $t_{\rm E}$ distribution, 
${\it \Phi}(t_{\rm E})$, can be calculated for an assumed mass function with 
a standard Galactic mass density and velocity model\cite{sumi03,alc00a,HanGould03}. 
We consider two mass functions. The first is a broken 
power-law\cite{HanGould03,Zoccali2000} d$N$/d$M=M^{-\alpha}$, with the power 
index of $\alpha_1 = 2.0$ for $0.7 \le M/M_\sun \le 1$, $\alpha_2 =1.3$ 
for $0.08 \le M/M_\sun \le 0.7$ and $\alpha_3$ as a fitting parameter for the 
brown dwarf regime $0.01 \le M/M_\sun \le 0.08$. The second is a log-normal 
function\cite{Chabrier2003} d$N$/d$\log M =\exp[(\log M-\log M_{\rm c})^2/(2\sigma_{\rm c}^2)]$ 
with a mean mass $M_{\rm c}$ and a width in $\log M$ of $\sigma_{\rm c}$, 
for $0 \le M/M_\sun \le 1.0$. For both mass functions, we assume that stars 
that were initially above 1 $M_\sun$ have evolved into stellar remnants --- 
white dwarfs, neutron stars or black holes, depending on their initial 
masses\cite{Gould2003} (see Fig. \ref{fig:tE} and Supplementary Table 3).

The mass functions were constrained by a likelihood analysis, with the 
likelihood function given by the product of the model probability 
${\it \Phi}(t_{\rm E})$ of finding $N_{\rm obs}= 474$ events with each of 
the observed $t_{{\rm E},i}$, that is; 
$L=\Pi_i^{N_{\rm obs}} {\it \Phi}(t_{{\rm E},i}) \epsilon(t_{{\rm E},i})$.

We have evaluated the likelihood distributions for these mass functions
both with and without the $t_{\rm E} < 2$ events, but the inclusion of 
the events with $t_{\rm E} < 2$ days makes little difference. The 
results are shown in Supplementary Table 3 and Supplementary Figs 6 and 7. 
Fig. \ref{fig:tE} indicates that both models match the data well for 
$t_{\rm E} \ge 2$ days, but at $t_{\rm E} < 2$ days, the ten observed 
events are well above the model predictions. The power-law and log-normal 
models predict 1.5 and 2.5 events with $t_{\rm E} < 2$ days, respectively, 
and the corresponding Poisson probabilities for the ten observed events 
are $4\times10^{-6}$ and $3\times10^{-4}$. Thus, we feel confident in 
adding a new planetary-mass population.

\begin{figure}
\includegraphics[angle=-90,scale=0.6,keepaspectratio]{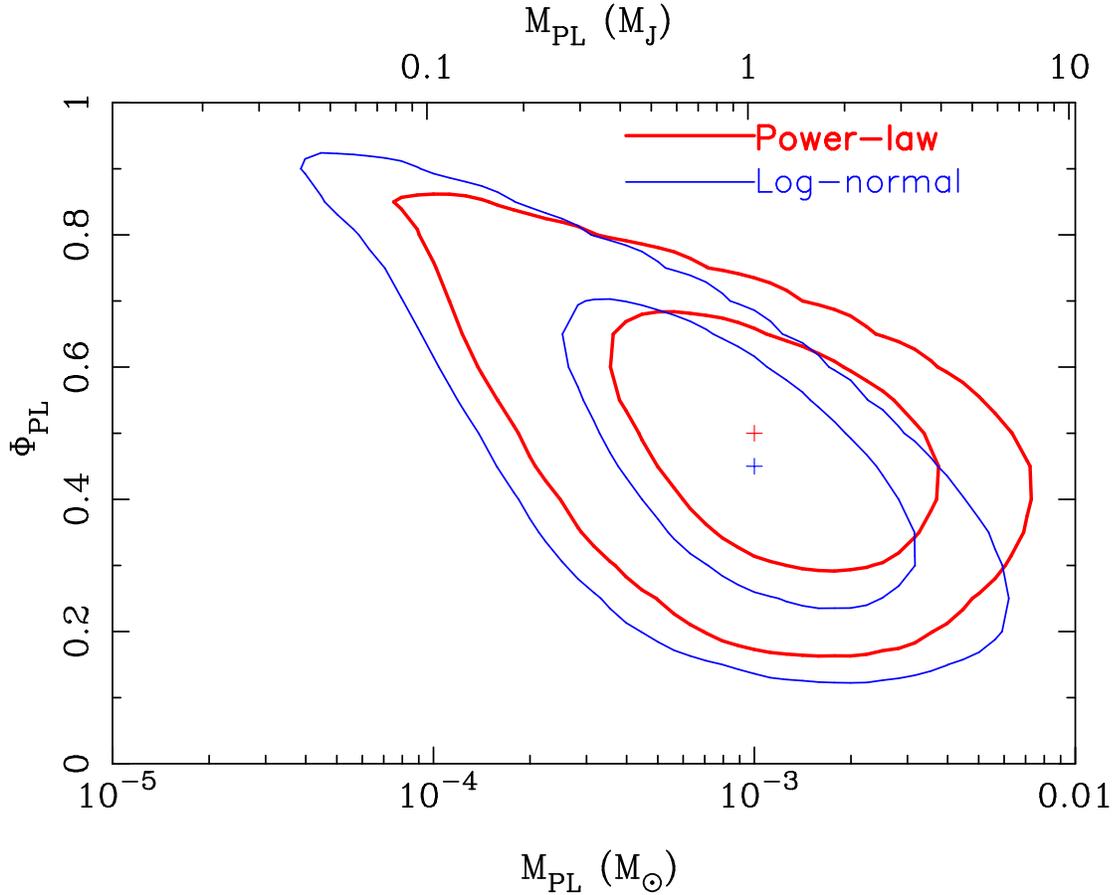}
\caption{
Likelihood contours for the planetary-mass function parameters. 
${\it \Phi}_{\rm PL}$ indicates the fraction of all objects in 
the planetary-mass population, not including the brown dwarfs 
that have planetary mass in the tail of the log-normal mass 
function. $M_{\rm PL}$ represents their mass. The two sets of 
contours indicate the 68\% and 95\% confidence levels. The red 
and blue curves indicate the power-law and log-normal mass 
functions, respectively, and "+" indicates the maximum likelihood 
points. The top axis scale is Jupiter masses, 
$MJ = 9.5\times 10^{-4} M_\sun$, while bottom axis scale is in 
Solar masses, $M_\sun$. For the power-law model, the likelihoods 
are evaluated in the ($\alpha_3$, $M_{\rm PL}$ , ${\it \Phi}_{\rm PL}$) 
space and projected into the ($M_{\rm PL}$, ${\it \Phi}_{\rm PL}$) plane. 
The $M_{\rm c} = 0.12$ and $\sigma_c= 0.76$ parameters are fixed 
for the log-normal model. The median and 68\% confidence intervals 
of ($M_{\rm PL}/M_\sun$, ${\it \Phi}_{\rm PL}$) are 
($1.1_{-0.6}^{+1.2}\times 10^{-3}$, $0.49_{-0.13}^{+0.13}$) and 
($0.83_{-0.51}^{+0.96}\times 10^{-3}$, $0.46_{-0.15}^{+0.17}$)
for the power-law and log-normal models, respectively. The results 
for two models are consistent with each other. The power-law and 
log-normal models imply $1.9_{-0.8}^{+1.3}$ and $1.8_{-0.8}^{+1.7}$ 
times as many unbound or distant Jupiter-mass objects as the main 
sequence stars. $\alpha_3$ is consistent with the values derived 
without planetary population indicating that brown dwarfs are 
$0.7\pm0.3$ times as common as main sequence stars. The numerical 
values of the models are summarized in supplementary Table 3.
\label{fig:likelihood}}
\end{figure}

For simplicity, we chose a planetary-mass function model with a 
$\delta$-function in mass $M_{\rm PL}$ and a fraction of all objects 
in the planetary-mass populations ${\it \Phi}_{\rm PL}$. The values of 
($M_{\rm PL}/M_\sun$¿,${\it \Phi}_{\rm PL}$) derived from the likelihood 
analysis are ($1.1_{-0.6}^{+1.2} \times 10^{-3}$, $0.49_{-0.13}^{+0.13}$) 
and ($0.83_{-0.51}^{+0.96} \times 10^{-3}$, $0.46_{-0.15}^{+0.17}$) for 
the power-law and log-normal models, respectively. The contours are shown 
in Fig. \ref{fig:likelihood}. Both models for ${\it \Phi}(t_{\rm E})$ 
provide good fits to the entire $t_{\rm E}$ distribution, as shown in 
Fig. \ref{fig:tE}. The power-law and log-normal models imply 
$1.9_{-0.8}^{+1.3}$ and $1.8_{-0.8}^{+1.7}$ times as many unbound or 
distant Jupiter-mass objects as main-sequence stars in the mass range 
$0.08 < M/M_\sun < 1.0$, respectively. These planetary-mass objects are 
at least 1.5 times as frequent as planets with host stars 
(see Supplementary Information). We tested a third mass function that 
has fewer massive stars and brown dwarfs\cite{Thies2007}, and found 
that the resultant planetary-mass function parameters are consistent 
with above values (see the Supplementary Information).

The lenses for these short events could be either free-floating planets 
or planets with wide separations of more than about ten astronomical units 
(AU) from their host stars, for which we cannot detect the host star in 
the light curves\cite{Han2003}. However, direct imaging, with adaptive 
optics, of planets orbiting young stars places upper limits on planets 
at wide separations. The Gemini Planet Imager has set upper 
limits\cite{Lafreniere2007} on the number of stars with Jupiter-mass 
planets at semi-major axes of 10-500 AU. From these results, we estimate 
that $<$0.4 of the 1.8 planetary-mass objects per star are likely to be 
bound to stars at orbital separations of $<$500 AU (see Supplementary 
Information Section 8). Hence, more than 75\% of these planetary mass 
objects are probably unbound to stars if their typical mass is a 
Jupiter-mass or more.

Since the $\delta$-function planetary models are not likely to be 
realistic, we also tested a fourth mass function that is identical 
to the first, broken-power-law model except for having a power-law 
form in the planetary mass regime, $M< 0.01M_\sun$. This yields a 
planetary-mass index of $\alpha_{\rm PL}= 1.3_{-0.4}^{+0.3}$, which 
is much steeper than the brown dwarf slope of $\alpha_3=0.49_{-0.27}^{+0.24}$,
indicating that they are distinct populations (see Supplementary Information).

Planet-formation theories predict that dynamical instabilities in 
planetary systems with multiple giant planets could scatter many of 
these planets into unbound orbits\cite{Vera2009}, as well as some 
into large separations\cite{Marois2008}. Recent observations also 
indicate that planet-planet scattering plays an important part in 
moving giant planets into short-period orbits\cite{Winn2010,Howard2010}. 
The planetary-mass population that we have identified here may have 
formed in protoplanetary disks at much smaller separations and then 
been scattered into unbound or very distant orbits.
\\
\\
\\
\\
{\bf Supplementary Information} accompanies the paper on www.nature.com/nature.
\\
\\
{\bf Acknowledgements} MOA thanks the support by the JSPS and MEXT of 
Japan, and the Marsden Fund of New Zealand. D.P.B. acknowledge supports 
by the NSF and NASA. The OGLE project is grateful for funding from the 
European Research Council Advanced Grants Program. 
\\
\\
{\bf Author contributions} T.S. and K.K. conducted data reduction 
and statistical analysis. A.U. produced OGLE-III light curves. 
I.A.B. generated the extended MOA-II light curves. D.P.B. 
conducted the detail analysis of binary events. T.S. wrote the 
manuscript. D.P.B. and I. A. B. edited the manuscript. All authors 
contributed to the observation and maintenance of observational 
facilities, discussed the results and commented on the manuscript.
\\
\\
{\bf Author Information} Reprint and permissions information is available at\\
npg.nature.com/reprintsandpermissions. The authors declare no competing 
financial interests. Correspondence and requests for materials should be 
addressed to T.S. (e-mail: sumi@ess.sci.osaka-u.ac.jp).

%-------------FIG.1--------------------

%-------------FIG.2-------------------- 

%-------------FIG.3--------------------

\begin{deluxetable}{lllrrrrrrcr}
\rotate\tablecaption{Microlensing Parameters of Short-Timescale Events.  \label{tbl:param}}
\tablewidth{0pt}
\tablehead{
\colhead{ID} & \colhead{field} & \colhead{R.A.} & \colhead{Dec.}  & \colhead{$N_{t_{\rm E}}$}& \colhead{$t_0$}          & \colhead{$t_{\rm E}$}    & \colhead{$u_{0}$} & \colhead{$A_{\rm max}$} & \colhead{$I_{\rm s}$} & \colhead{$d_{\rm min}$} \\
\colhead{} & \colhead{} &  \colhead{(2000)}  & \colhead{(2000)} & \colhead{} & \colhead{(JD$'$)}         & \colhead{(days)}           & \colhead{$(R_{\rm E})$}  & \colhead{}       &  \colhead{(mag)} & \colhead{$(R_{\rm E*})$}}
\startdata
     MOA-ip-1 &  gb1-4 & 17:46:24.506 & -34:30:36.82 &   9 & 3883.24171 & 0.73 $\pm$ 0.08 & 0.028 $\pm$ 0.003 & 35.6 & 19.7 & 7.0\\
     MOA-ip-2 &  gb4-3 & 17:52:34.143 & -30:54:14.25 &  28 & 4223.88851 & 0.49 $\pm$ 0.10 & 0.400 $\pm$ 0.212 & 2.6 & 17.9 & 3.3\\     
     MOA-ip-3 &  gb5-7 & 17:54:58.325 & -29:38:20.68 & 170 & 4295.34720 & 1.88 $\pm$ 0.12 & 0.911 $\pm$ 0.096 & 1.4 & 17.2 & 3.6\\
     MOA-ip-4 &  gb5-8 & 17:54:24.543 & -29:13:29.39 &  81 & 3961.38803 & 1.48 $\pm$ 0.12 & 0.271 $\pm$ 0.061 & 3.8 & 19.2 & 3.1\\
     MOA-ip-5 &  gb9-2 & 17:57:17.008 & -29:02:33.59 &  69 & 4169.60907 & 1.62 $\pm$ 0.69 & 0.126 $\pm$ 0.159 & 8.0 & 19.2 & 2.4\\
     MOA-ip-6 &  gb9-4 & 17:59:19.977 & -29:31:24.70 &  27 & 4189.49214 & 1.78 $\pm$ 0.24 & 0.499 $\pm$ 0.122 & 2.2 & 18.3 & 4.8\\
     MOA-ip-7 &  gb9-5 & 17:57:36.678 & -29:59:40.52 &  51 & 4370.69496 & 1.82 $\pm$ 0.87 & 0.143 $\pm$ 0.125 & 7.0 & 19.4 & 5.2\\
     MOA-ip-8 &  gb9-5 & 17:59:34.877 & -30:04:24.04 &  47 & 4013.14052 & 1.36 $\pm$ 0.15 & 0.103 $\pm$ 0.016 & 9.8 & 18.8 & 4.8\\
     MOA-ip-9 & gb10-5 & 17:57:52.952 & -28:16:56.66 &  16 & 3910.81772 & 0.96 $\pm$ 0.21 & 0.163 $\pm$ 0.058 & 6.2 & 19.5 & 3.4\\
    MOA-ip-10 & gb11-9 & 18:09:00.076 & -32:18:39.91 &  21 & 3932.99205 & 1.19 $\pm$ 0.04 & 0.032 $\pm$ 0.001 & 30.8 & 18.8 & 15.0\\
\enddata
\tablecomments{
$N_{t_{\rm E}}$ indicates the number of data points within $t_0\pm t_{\rm E}$ 
and $t_0$, $t_{\rm E}$, $A_{\rm max}$ and $I_{\rm s}$ indicate the time of 
peak magnification, the Einstein radius crossing time, the maximum magnification, 
and the source star magnitude of the best fit models of the MOA data, 
respectively. JD$'$=JD-2450000. $u_0$ and $d_{\rm min}$ indicate the 
source¿lens impact parameter and minimum host star separation in units of 
the Einstein radii of the planetary mass lens, $R_{\rm E}$, and 
possible host star, $R_{\rm E*}$, respectively. The errors in $t_{\rm E}$ 
and $u_0$ represent 1-$\sigma$ limits. $d_{\rm min}$ indicates the 2-$\sigma$ 
limits. MOA-ip-2, MOA-ip-3 and MOA-ip-10 were alerted as MOA-2007-BLG-144, 
MOA-2007-BLG-309 and MOA-2006-BLG-098 by the MOA real-time alert system 
(http://www.massey.ac.nz/~iabond/moa).
}
\end{deluxetable}

\clearpage

\section*{\center Supplementary Information}
\hspace{1.5em}
This supplement to ``Unbound or Distant Planetary 
Mass Population Detected by Gravitational Microlensing\rlap," by Sumi et al.\ provides 
additional details of the analysis presented in this paper.
The basics of gravitational microlensing is briefly 
introduced in Section \ref{sec:microlensing}.
The detailed data analysis and event selection methods
are described in Section \ref{sec:analysis}. In Section \ref{sec:Background},
we discuss possible backgrounds that could conceivably
result in false detections of short timescale microlensing event, and
we show that the short timescale event sample does not have
any significant contamination by false-positive detections.
We discuss the simulations used for the detection efficiency 
calculations in Section \ref{sec:simulation}. In Section \ref{sec:systematic},
we show that systematic biases of the parameters produced
by our analysis are small, and in Section \ref{sec:Likelihood}, we
present the details of the likelihood analysis that measures the
substellar mass function.  The likelihood analysis for other mass 
functions are also given in Section \ref{sec:Likelihood2}.
Finally, in Section \ref{sec:Frequency}
we compare the frequency of unbound or distant planetary mass objects 
that of bound planets.

%---------------------------------------------

%\clearpage
\section{Basic of the gravitational microlensing}
\label{sec:microlensing}
 
\hspace{1.5em}
The gravitational microlensing method allows the detection of extremely 
faint or completely dark objects when their gravitational fields act as 
a lens to magnify background source stars\citep{pac86}.
A gravitational lens is characterized by its Einstein ring radius, 
which is the radius of the ring image seen when the lens and source 
are perfectly aligned. In a microlensing event, the image separation 
is too small to resolve, and the observable feature is the changing 
total magnification due to the lens-source relative motion. The key 
parameter of a microlensing light curve is the Einstein radius crossing time
given by
\begin{equation}
  \label{eq:lens_eq}
 t_{\rm E}=\frac{R_{\rm E}(M,D_{\rm l},D_{\rm s})}{v_{\rm t}}, \ {\rm where}\ \ R_{\rm E}=\sqrt{\frac{4GM}{c^2}\frac{D_{\rm l}(D_{\rm s}-D_{\rm l})}{D_{\rm s}}},
\end{equation}
where $v_{\rm t}$ is the relative transverse velocity between the lens and source,
$R_{\rm E}$ is the Einstein radius, $M$ is the lens mass, $D_{\rm l}$ and  $D_{\rm s}$ are
the distance to the lens and the source from the observer, $G$ is the gravitational 
constant and $c$ is the speed of light. 
Although $t_{\rm E}$ depends on the velocity and distance
of the lens object, in addition to its mass, the $t_{\rm E}$ distribution can
be used to probe the mass function of the lens objects statistically.
Thousands of microlensing events due to stars in the Galactic disk 
and bulge have been detected and used to study the stellar mass function 
and galactic structure \citep{uda94,alc00a,sumi03,Hamadache2006}.

\section{Data Analysis and Event Selection}
\label{sec:analysis}
\hspace{1.5em}
The 2006-2007 MOA-II galactic bulge data consists of $\sim 8250$ images
of each of the two most densely sampled fields (fields gb5 and gb9) and
1660-2980 images of each of the 20 other less densely sampled fields
as shown in Table \ref{tbl:fields}. We observed these galactic bulge 
fields for about 8 months of each year from the end of 
February to the beginning of November. This observing strategy is designed to
detect single lens events and short timescale 
anomalies in the light curves of stellar microlensing events
due to planets orbiting the lens stars
\citep{mao1991,Gould2010,Sumi2010}.
The MOA images were reduced with MOA's implementation \citep{bon01}
of the difference image analysis
(DIA) method \citep{tom96,ala98,ala00}. In the DIA method, a high quality, 
good seeing, reference image is subtracted from each observed 
image after transforming the reference image to give it the same seeing
and photometric scaling as the observed image. This method generally 
provides more precise photometry in the
very crowded Galactic bulge fields 
than straight PSF-fitting routines, such as DOPHOT \citep{dophot}.
This is, in part, due to the fact that the Galactic bulge fields are
so crowded that virtually all the main sequence stars 
are not individually resolved. As described in Section \ref{sec:simulation},
the identification of a clear red clump giant population in the data
is needed to match the observed MOA luminosity function to the much
deeper Hubble Space Telescope luminosity function that describes the
source stars. For the gb22 field and some fraction of other fields, totaling 
about 12\% of the area, a clear red clump population could not
be identified in the color magnitude diagram, and these regions
were excluded from the analysis for this reason. This lack of a
clear red clump giant feature in field gb22 is probably because it
is relatively far from the center of the Galaxy. It is generally regions of
very high interstellar extinction that prevented the identification of the
red clump giant color magnitude diagram feature in portions of the
other fields.

The images were taken using the custom MOA-Red wide-band filter, which
is equivalent with the sum of the standard Kron/Cousins $R$ and $I$-bands.
The instrumental magnitudes of the MOA reference images were
calibrated to the Kron/Cousins $I$-band using OGLE-II photometry map of the 
Galactic bulge\citep{uda02}.
The mean magnitude zero-point were estimated from the 30\% of MOA-II fields which 
overlap with the OGLE-II map. We applied this mean zero-point to all fields.
Here the error of the calibrated magnitude is estimated to be $\sim$0.25 mag
from the standard deviation of zero-points in overlap fields.
Although this calibration is approximate, this does not affect 
following analysis at all because the luminosity functions, which is the
only part of our analysis requiring calibrated magnitudes, are
calibrated by using red clump giants, as discussed in Section \ref{sec:simulation}.

The problem of microlensing event selection has been addressed in
previous microlensing analyses 
\citep{alc00a,alc00b,sumi03,pop05,Hamadache2006,Sumi2006,Wyrzykowski2009},
and we adopt event selection criteria that are similar to those
used in a number of these previous papers. We employ three levels
of selection criteria, or cuts, to distinguish 
genuine microlensing events from intrinsic variable stars
and astrophysical and non-astrophysical artifacts.

\begin{itemize}
\item[$(1)$]
Cut-0:
On a subtracted image, variable objects can be seen as positive or 
negative point spread function (PSF)
profiles depending on whether the target star is brighter
or fainter than in the reference image. We use a custom implementation
of the IRAF task DAOFIND \citep{daophot} to
detect these variable objects (like microlensing events), with the modification
that both positive and negative PSF profiles are searched for simultaneously. 
This algorithm finds difference image peaks with a signal to 
noise ratio (S/N) $> 5$ and then
applies several additional criteria to avoid the detection of 
spurious variations that are not associated with stellar variability,
such as cosmic ray hits, satellite 
tracks and electrons leaked from the saturated images of bright  stars. 
The positions of detected objects are checked against those obtained in previous 
reductions of the field.  When no object is cross-referenced, the object is
classified as new and added to the list of variable object positions.
If an object has previously been detected, 
the number of detections for this object, $N_{\rm detect}$, is 
incremented. We require that $N_{\rm detect}$ to be $> 2$.
About $9.7\times10^6$ variable objects were detected at this stage
of the analysis, including a number of image artifacts of various types.

\item[(2)]
Cut-1:
Light curves of the candidates passing Cut-0 were then created using
PSF fitting photometry on the difference images. Only objects with more
than 500 photometric data points (from 1660-8250 images depending on the field)
were used for further analysis. The photometric error bars 
were calibrated with constant stars in each sub-field.
We place a 120-day moving window on each
light curve and define the baseline flux to be the weighted average flux
outside of that window. 
We require the baseline to have more than
10 data points and have $\chi_{\rm out}^2/{\rm dof} \le 3$ for a constant flux fit.
As other
microlensing surveys have found, the error bar estimates from the photometry
code provide only an approximate description of the photometric uncertainty for
each measurement. The actual uncertainties depend upon the distribution of
stars in the immediate vicinity of the target.
We therefore define normalized uncertainties
$\sigma'_i \equiv \sigma_i \sqrt{\chi^2_{\rm out}/{\rm dof} }$,
where $\sigma_i$ are the error bar of the measured flux $F_i$ for the $i$th
measurement.
We then search for positive light curve ``bumps" inside
the 120-day window, with a ``bump" defined as a brightening episode with 
more than 3 consecutive measurements with excess flux $> 3\,\sigma'$
above the baseline flux, $F_{\rm base}$.
We define a statistic $\chi_{3+} = \Sigma_i \left( F_i - F_{\rm base}  
\right)/\sigma'_i$
summed over consecutive points with $F_i - F_{\rm base} > 3\sigma'_i$.
We require that a bump should have $\chi_{3+} \geq 80$ to pass this cut.

\item[$(3)$]
Cut-2:
The candidate events that pass Cut-1 are fit with a 5-parameter microlensing
model assuming a point source and a single lens object\citep{pac86}.
The 5 parameters are the Einstein radius crossing time, $t_{\rm E}$, the time
of peak magnification, $t_0$, the source-lens impact parameter (in units of
the Einstein radius), $u_0$, and the source and background fluxes, $F_{\rm s}$ and
$F_{\rm b}$. 
Candidate events are rejected when the Markov Chain Monte Carlo (MCMC)
fitting method does not converge, because such events generally do not
have well constrained parameters and may add a systematic bias in the observed
$t_{\rm E}$ distribution (see Section \ref{sec:systematic} for more details). 
We place constraints on the fit $\chi^2$ 
for these microlensing fits. We require that $\chi^2/{\rm dof} < 2$ for both the
full fit and for the regions near the peak with $|t-t_0|< t_{\rm E}$ and $|t-t_0|< 2t_{\rm E}$.
We require that $t_0$ be within the observation 
period, and we exclude very low-magnification events by requiring $u_0 \le 1$
with an error bar of $\sigma_{u_0} < 0.3$.
We also limit the best fit Einstein radius crossing times to the range
$ 0.3\,{\rm days} \le t_{\rm E} \le 200\,$days with an error bar of $\leq 50$\% 
and $\leq 12$ days, and we require that the
source should have an (approximately calibrated) $I$-band magnitude brighter
than 20.0. We require that the best fit source flux not exceed the 
cataloged flux on of an apparent source on the reference frame by more than
a factor of 3. This allows for some ``negative blending", which can occur if
a source happens to be located at a local minima in the flux of unresolved stars.

Of course, most of the cuts on the fit parameters exclude legitimate microlensing
events, but we use them because they allow us to exclude the background
of non-microlensing events with a relatively simple set of cuts. The cuts
on $t_{\rm E}$ are of particular interest because these will effect the shape of the
$t_{\rm E}$ distribution that is one of the major results of this analysis. The
$t_{\rm E} < 200\,$day cut (and the use of a 120-day window in Cut-0) excludes a
large number of long events, which would be very important if we were trying
to measure the high-mass end of the present-day mass function. This is 
not a problem for this analysis, which focuses on the low-mass end of the mass
function. The removal of short events with $t_{\rm E} < 0.3\,$days probably does not
remove any events from this sample, but there are certainly real events
of such short timescales. It will be important to remove this cut in future
analyses of larger event samples that will measure the shape of the 
mass function of the unbound or distant planetary mass objects.

Events with systematic residuals from the best fit model are also rejected, and this
cut depends on the significance of the microlensing signal.
We define the maximum number of consecutive measurements that are scattered 
from the best fit model with excess flux more than 2-$\sigma$ and 3-$\sigma$ 
by $N_{2\sigma}$ and $N_{3\sigma}$, respectively. We require 
that $\chi_{3+} \ge 70 N_{2\sigma}$ -500, and 
$\chi_{3+} \ge 45 N_{3\sigma}$ or $N_{3\sigma} \le 2 $.

\end{itemize}

All event selection criteria are summarized in Table \ref{tbl:criteria}. We have
selected events which can be well fit with a point-source, single-lens model,
and we have omitted binary lens events in this analysis by the cuts above, although
some previously unknown planetary-mass binary lens events have been 
found in this analysis (Bennett et al. in preparation).
Although we have identified more than a thousand microlensing candidates 
in this data set, only 474 high quality microlensing events have passed our 
relatively strict criteria. These strict criteria ensure that $t_{\rm E}$ is well constrained
for each event and that there is no significant contamination by 
mis-classified events, which might bias the $t_{\rm E}$ distribution. This
potential contamination by mis-classified events is discussed
in the next Section.

\section{Background Events}
\label{sec:Background}
\hspace{1.5em}
There are a variety of effects that can give signals that bear some resemblance 
to short timescale microlensing events, so it is important to ensure that
our sample of events is not contaminated by any of these effects. 
The potential background effects that we have considered are the following:

\begin{itemize}
\item[(1)]
Cosmic-ray hits: 
It is rare for cosmic ray hits to give a signal with the same profile
as the observed PSF. So, most cosmic rays are rejected due to
their lack of a PSF-like shape. There was a previous claim of the 
detection of free-floating planetary mass objects by microlensing in 
the globular cluster M22 based on imaging by the 
{\it Hubble Space Telescope} (HST) \citep{Sahu2001}, but this claim was
retracted within a few months \citep{Sahu2002} as the putative 
microlensing events were artifacts caused by cosmic ray hits 
very close to the targets stars on two consecutive images. Our
data set is not subject to such confusion for a number of reasons.
First, unlike the MOA images, HST images are under-sampled, so that 
cosmic rays are much more difficult to reject based on their PSF shape in
the HST images than in the MOA images. Second, the light curves
presented in the M22 paper are maximally under-sampled, with only
a pair of consecutive (cosmic-ray split) images that were brighter than
the baseline brightness. In contrast, we explicitly require at least
4 consecutive observations that are $> 3\,\sigma'$ above baseline
and implicitly require much larger number with our cuts on the fit
parameter error bars. Finally, the cosmic ray rate is much lower for
ground-based images than for HST images. In fact, each of our
10 $t_{\rm E} < 2\,$day events have at least 10 observations at
significant magnification, so there is no chance of contamination by
cosmic ray hits.

\item[(2)]
Fast Moving objects: Our pipeline conducts PSF photometry at the fixed 
position where the variable object was first detected on a difference image. 
If an object moves a significant fraction of PSF during several exposures, 
the photometry code will begin to underestimate its brightness due to
the off-center PSF. With this photometry method, the linear 
movement of such an object can produce a light curve
which has some resemblance to a
microlensing light curve.  Any small Solar System body, such as an
asteroid or Kuiper belt object,
with a proper motion of $\sim 1 {\rm asec}\,h^{-1}$, can mimic a
short timescale microlensing event. Dust specs on the camera window 
can also produce similar light curves due to slight changes in
the telescope pointing between different images of the same field.
While light curves due to moving objects or dust specs can sometimes
be reasonably well fit with a microlensing light curve model, they
generally give unphysical light curve parameters, such as
a large $u_0$ value, a small $t_{\rm E}$ value, and a fit source
brightness, which to bright to be compatible with stars seen in the
un-subtracted images. 
The moving objects in this data sets were all rejected by two independent 
methods. We have identified 3743 moving objects in this data set
by direct measurements
of the changing image positions on the difference images, and we found
that all the light curves of these objects were rejected due to poor 
microlensing fits or unphysical microlensing model parameters.

\item[(3)]
Cataclysmic variables (CVs): 
The most important background for this analysis is a particular class
of CVs that can have short timescale brightening episodes and repeat
rarely enough so that they might have been missed despite the 
multi-year, high cadence monitoring of the MOA-II bulge fields.
Our primary means for identifying CVs is through the microlensing
light curve fits. Most of the CVs can be identified from their highly
asymmetric light curves, with a steep rise and a slow decline. However,
for some CVs, this asymmetry is not seen due to gaps in light curve
sampling or large photometric error bars. However, even poorly sampled 
CVs can generally be distinguished from microlensing light curves because 
a theoretical microlensing profile fit yields unphysical values such as 
a very large value of $u_0$ and a sources with much brighter baseline 
fluxes than allowed by the reference images

We identified 418 CVs in our sample from a visual inspection of light
curves that failed the cuts on the quality of the microlensing fit or the upper
limit on the brightness of the source star. The timescales of the brightening
of these CVs range from hours to months. An important property of CVs
is that they repeat \citep{CVs},  and we have used this property to
find an upper limit on the CV contamination of our sample. Using an
expanded data set running through 2010, we found 208 CVs with more
than one outburst. The average interval between microlensing events with the
same source is about 100 thousand years if the lens objects are
not physically related. 
The rate of repeating microlensing events due to wide 
binary lens or source systems is $<1$ in a few hundred events, so 
very few, if any, of these events are likely to be actual microlensing events.
Slight contamination of this CV sample with microlensing events is 
not a problem 
because these CV samples are only used for determining an
upper limit of CVs which could pass our selection criteria.

This upper limit is derived as follows.
Because these CVs were identified by their multiple outbursts instead of
the light curve shape, they can be used as a background sample to check 
how well we can reject CVs by their light-curve shapes. We decomposed the light 
curves of these multi-outburst CVs into 421 individual outbursts. 
The light curves of these outbursts were fit with microlensing 
models, and we found that none of them passed our selection criteria within $0.3<t_{\rm E}<2$ days. 
With zero false positive in 421 outbursts, Poisson statistics provides an
upper limit of 2.3 at the 90\% c.l..  Then, the expected 90\% c.l. 
upper limit of contamination among the 197 single-outburst CVs in our 
sample is 2.3$\times$197/421=1.1, or 11\% of our ten 
$t_{\rm E}< 2\,$day events.

After the MOA event selection was complete, we extended the light curves
by adding the 2008-2010 MOA data 
and found that none of the ten $t_{\rm E} < 2\,$day events 
have any other brightening episode. The MOA group also requested the OGLE
light curves of these events. Seven of the ten short events were also observed 
by OGLE-III, the third phase of OGLE survey, and none of them have any other 
brightening in the 8-year OGLE-III light curves. Furthermore, the event MOA-ip-3 
was also observed during OGLE-I and OGLE-II surveys and exhibits no further 
variations, which gives 16 years of coverage.
Although the event MOA-ip-10 was not observed by OGLE-III, it was observed 
during OGLE-II survey in 1996-2000 and no brightening was seen.
Finally, we checked the public database of the MACHO project, and 
found that the light curves of MOA-ip-3, MOA-ip-5, MOA-ip-6, and
MOA-ip-8 had light curves
with no photometric variation observed during the 1993-1999 observing seasons.
These data support our conclusion that these short events are not CVs.

\item[(4)]
Background supernovae:
The dominant background in the microlensing survey toward
the Large and Small Magellanic Clouds (LMC and SMC) was supernovae
in galaxies at large distances behind the Magellanic Clouds,
and methods to remove this 
background from the microlensing sample by means
of their asymmetric
light curves were developed \citep{alc00b,Rest2003,Tisserand2007}. 

The event rate of background supernovae in galactic bulge microlensing surveys
is much smaller than that toward the Magellanic Clouds for the following reasons.
The interstellar extinction toward the galactic bulge is much higher, 
$A_V=1-6$ mag \cite{sumEX04}, than towards the Magellanic Clouds
with $A_V\sim0.3$ mag \cite{Bessell1991}. Also,
the LMC is $\sim 6.25$ times more distant than the galactic bulge, and the SMC
is even more distant, so the Magellanic Could source stars tend to be
$\sim 2$ mag fainter. The ratio of supernovae to microlensing events is also
dramatically reduced by the factor of $\sim 100$ ratio between the high
bulge microlensing rate \cite{alc00a,pop05,Hamadache2006,Sumi2006} and
the low rate towards the Magellanic 
Clouds \cite{alc00b,Tisserand2007,Wyrzykowski2009}.

Although supernovae do not repeat like CVs, 
the shape of supernova light curves is somewhat similar to those of CVs.
Both tend to be highly
asymmetric with a steep rise and a slow decline.
So that they can be rejected by the microlensing fits\citep{alc00b,Rest2003,Tisserand2007}
in the same way as we have rejected CVs. 

Although supernovae do not pass our microlensing selection criteria, they could not
contaminate our short timescale event sample even if they did, as they have timescales
of $\sim$30 days or longer.

\item[(5)]
Binary microlensing events:
Binary lens systems generate small caustics when the separation of the
lens masses is either much smaller or much larger than the Einstein radius.
Wide binary events, however, can only mimic a short timescale single lens
event when one of the lens masses has a low mass \citep{DiStefano1999a,DiStefano1999b}. 
So, wide binaries are
not really a background, and we consider them in \S~\ref{sec:Frequency} where
we consider the distinction between very wide and free-floating planets.
The 2-$\sigma$ limits on the minimum separation to a host star are shown 
in Table $1$ in the main paper.
These limits are independent of the mass ratio and $R_{\rm E}$ value for 
the host star in the limit where $q \ll 1$.  

The situation with close binaries is different, because close 
binaries have two small caustics far from the center of mass that can give rise
short duration lensing events even when the masses of the lens
objects are large. We have fit all ten of the $t_{\rm E} < 2\,$day events with
close binary models and found that for most of the short events, a close-binary
model is strongly excluded, being disfavored over the single-lens model
by $\Delta\chi^2 \geq 30$. The only exceptions are events with relatively
poor sampling. For events MOA-ip-2 and  MOA-ip-6, the close binary models
are disfavored by $\Delta\chi^2 = 17.2$ and 9.8, respectively. These imply
formal probabilities of $<0.02\,$\% and $< 0.8\,$\% for these events to
be caused by close binaries.

Event MOA-ip-5 is the only event for which the close binary model cannot
be rejected. The best fit model for this event is actually a planetary wide binary
model, which has a $\chi^2$ value lower than the single lens $\chi^2$ value
by $\Delta\chi^2 = 27.8$. However, this level of $\chi^2$ improvement is
likely to be due to systematic photometry errors spread over a large number
of data points, and we do not consider this a large enough $\chi^2$
improvement to indicate the detection of a host star. The best close binary
model also gets some $\chi^2$ improvement from regions of the light
curve that are distant from the peak shown in Fig. \ref{fig:lc}, and
it has a $\chi^2$ value that is better than the best fit single lens light
curve by $\Delta\chi^2 = 11.9$ with 4 more fit parameters (as the 
magnification of the single lens model is too low for finite source
effects to be important). Formally, this would be significant at the
98\% confidence level, but only $\Delta\chi^2 = 5.5$ of this difference
comes from the region within 10 days of the light curve peak. A
$\Delta\chi^2 = 5.5$ difference is only significant at the $\sim 70$\% 
confidence level, so we do not consider the close binary model
to be be significantly favored for this event. The light curve 
coverage is insufficient to distinguish 
between the $t_{\rm E} < 2\,$day single lens and close binary models. 

We can compare the case of MOA-ip-5 to 
the short events with better sampling in order to estimate the probability that
MOA-ip-5 is due to a close binary lens system and is not a short timescale
event. This analysis has revealed the 9 other events listed in 
Table $1$ in the main paper, 2 wide separation planetary 
(or brown dwarf) binary events 
and one close binary (Bennett et al. in preparation). There is
one other short event that is a poor
fit to a single lens event, but a worse fit to a close binary. (It is probably
a CV.) Thus, there are $\sim 11$ short events that are best fit by
single lens or wide binary models, but only one that are best fit
by a close binary. So, it is unlikely that MOA-ip-5 is a close binary
event. The probability that it is a close binary can be quantified in
an analysis like that by \citet{bennett05}, but it is simplest to just
assume that MOA-ip-5 is a single lens event.

We can also constrain the short event contamination due to close binaries
by using the observed binary fraction and distribution.  About 30\% of main 
sequence stars have binary companions \citep{Lada2006}. Only the small caustics 
due to close binaries with separation $0.05<d/R_{\rm E}<0.3$ in Einstein radius 
unit can resemble the short single-lens microlensing events.
For $0.05>d/R_{\rm E}$, the caustics become too small to be detected.
For $d/R_{\rm E}>0.3$, the effect of the host star is observable.
This range corresponds $-1<\log (a/{\rm AU})<0$ for the typical lens 
systems, i.e., K or M-dwarfs. Only 6\% of binaries have separation within this 
range \citep{Duquennoy1991,Delfosse2004}. The event probability and 
detection efficiency for short events, which depend only on magnification 
timescale, are equivalent for close binary events and for the single
lens events due to unbound or distant planetary mass objects. So, we can 
directly compare the number of close binary and single lens systems 
to estimate the contamination by close binaries. The 
fraction of short events due to close binaries versus
single lens events is $0.3\times0.06\times2/1.8=0.02$. There is an
additional factor of two for the close binaries because 
of the two small caustics generated 
a close binary lens system. This is consistent with our conclusion
above that the contamination of our sample of short events by
close binary lens systems is negligible.

\item[(6)]
Microlensing by high velocity stars and Galactic halo stellar remnants.
Stars with very high velocity could also produce short timescale events.
The mean timescale of the events due to stars in the bulge is
$t_{\rm E}=R_{\rm E}/v_{\rm t}=R_{\rm E}/(D_{\rm l}\mu)\sim 20\,$days, where $\mu=v_{\rm t}/D_{\rm l}$ is
the lens-source relative proper motion. So, it would require stars
with at least an order of magnitude higher velocity than the typical
stars to explain the events with $t_{\rm E} \le 2\,$days.

For lens stars with high velocity at fixed distance $D_{\rm l}$, 
$t_{\rm E}$ is proportional to $\mu^{-1}$. The typical proper motion of stars 
in our bulge fields is $\mu\sim$6 mas\,yr$^{-1}$, so lens stars 
with $\mu \ge 60$ mas\,yr$^{-1}$ can have $t_{\rm E} \le$2 days. 
However the number of such high proper motion stars with 
$\mu \ge 60$ mas\,yr$^{-1}$ are only $6\times10^{-5}$ of stars in the 
OGLE-II proper motion catalog in the galactic bulge fields \citep{sumi04}. 
The event rate per lens star is proportional to the Einstein 
radius, $R_{\rm E}$, times a transverse velocity, $v_{\rm t}$, so the event rate due to 
the high proper motion stars with a typical mass of 0.3\,$M_\sun$ is a 
factor of $\sqrt{0.3M_\sun}/\sqrt{0.001M_\sun}\times60/6=170$ higher 
than that of a Jupiter mass lens with typical kinematics.  If we make the
(unrealistic) assumption that all the high proper motion stars are high
velocity stars in the bulge, then we can
obtain the expected rate of $t_{\rm E} \le 2\,$day events due to these
high velocity stars. This is simply the fraction of high proper motion
stars times the difference in event rate, or 
$6\times10^{-5} \times 170=0.01$ per star. This compares directly to the
inferred number, 1.8, of Jupiter mass lens objects from the main 
analysis. So, the background of high velocity stellar lenses is more than two
orders of magnitude less than the signal.
Moreover, a more straight forward interpretation of these high proper 
motion stars in the OGLE-II proper motion catalog is that they
are just just nearby stars instead of high velocity stars. Microlensing
events due to nearby high proper motion stars are already included
in our simulations of the $t_{\rm E}$ distribution, so they do not produce the
observed population of short $t_{\rm E}$ events.
Thus, the above upper limit of 0.01 per star is a very conservative limit.

The conclusion that these high proper motion stars are 
nearby stars with regular kinematics instead of
 more distant high velocity stars are the following: 

(a) The proper motion distribution in the OGLE-II catalog of
the bulge follows a slope of $\log(N)=-3\log(\mu)+{\rm const.}$, 
without any distinct features, for $\mu > 10$ mas\,yr$^{-1}$. 
This matches the expectation for a standard kinematical model and a
uniform distribution of stars in space \citep{sumi04}. This implies
that most of these high proper motion stars are 
not high velocity stars, but are instead nearby stars within 
about $300$\,pc which correspond the scale height of the disk.
The proper motion of disk stars at a distance of $<300$\,pc with 
velocity of $v\sim 30\,{\rm km\,s}^{-1}$ is $>20$\,mas\,yr$^{-1}$.

(b) Galactic bulge lens stars contribute about 70\% of all events observed
towards the bulge,  and the typical bulge star velocity is 
about $120$\,km\,s$^{-1}$. A high velocity star, with a velocity 
an order of magnitude higher than this, would have a velocity 
higher than the escape velocity from the Galaxy.
Galactic bulge radial velocity observations
found that no stars with radial velocity of $v_{\rm l}>400$\,km\,s$^{-1}$ 
in a sample of 3200 stars and  no significant deviations from a Gaussian 
distribution \cite{Howard2008}. This confirms that there is no high 
velocity population in the bulge that could explain these short events.

These arguments indicate that there is no significant population of 
high velocity stars to explain our detection of short events.

A Galactic halo population of 
ancient stellar remnants, such as white dwarfs, neutron stars or 
black holes have sometimes been suggested as a explanation of the
microlensing events seen towards the LMC by the MACHO
Collaboration\cite{alc00b}. If so, their velocities should be
$v\sim$200\,km\,s$^{-1}$.
The typical $t_{\rm E}$ value due to the stellar remnants in the halo is 
$> 20\,$days, so to get events with $t_{\rm E} \le 2\,$days, 
the velocities must be an order of magnitude 
larger, $v\sim$2000\, km\,s$^{-1}$. So, like the stars
considered above, a Galactic halo population of stellar remnants cannot
explain these short events due to the Galactic escape velocity
constraint.

A final test of the possibility of high velocity lenses could be
made by the detection of finite source effects for short events
with relatively high magnification. Microlensing events due to
massive lenses (with large $R_{\rm E}$) and  high velocities 
should show this effect much less frequently than events due to
planetary mass lenses (with small $R_{\rm E}$) and more typical velocities.
In the current sample, events ip-1 and ip-10, have magnification
high enough so that they might possibly show finite source effects.
A finite source model for event ip-1, yields an improvement of 
$\Delta\chi^2 = 1.6$ for a finite source model, which would imply
a relative proper motion of $\mu \sim 6\,$mas/yr, which is quite 
typical of bulge lensing events. For event ip-10, the best finite
source model improves $\chi^2$ by only $\Delta\chi^2 = 0.1$, but
the implied lower limit on the relative proper motion is only
$\mu > 4\,$mas/yr, so most of the range of expected values is
allowed.

We can expand this test with two short events from the MOA alert system
that occurred too late to be in our statistical sample. Event
MOA-2009-BLG-450 has $t_{\rm E} = 1.2\,$days and a peak magnification
of 85. It shows no finite source effects, which implies a relative 
proper motion lower limits of $\mu > 6\,$mas/yr. MOA-2010-BLG-418
was one of the shortest events known, with $t_{\rm E} = 0.41\,$days and
a peak magnification of 23. It does show a 3-$\sigma$ signal for a
finite source effect corresponding to a relative proper motion of
$\mu = 7\,$mas/yr. Thus, the event most sensitive to finite source
effects does show them, and the other events have limits on
$\mu$ which are consistent with expectations for a planetary 
mass population.

\end{itemize}

\section{Detection Efficiency Simulations}
\label{sec:simulation}
\hspace{1.5em}
To compare observed $t_{\rm E}$ distribution with the model, $\Phi_{t_{\rm E}}$,
we estimated the detection efficiency of our experiment using 
Monte Carlo simulations following previous work in the field\cite{sumi03}.
Artificial microlensing events were added at random positions in the 
observed images, using PSFs derived from nearby stars in each field.
The parameters of these artificial events were uniformly generated 
at random in the following ranges for the impact parameter, $u_0$,
time of peak magnification, $t_0$, Einstein radius crossing time,
$t_{\rm E}$, and source magnitude, $I$: $0 \le u_0\le 1.5$, 
$2453824 \le t_0\le 2454420$ JD, $0.1 \le t_{\rm E} \le 250\,$days, and
$14.25\le I \le 21.15$ mag. (The $t_0$ range is
the range of observations in this data set.)
The source magnitudes were weighted by the combined Luminosity function (LF)
from MOA and the {\it Hubble Space Telescope} (HST)\cite{hol98}. This uses
the MOA LF for bright stars and HST for faint stars down to $I = 24$.
This combined LF is calibrated to the extinction Galactic bulge distance for
each field using the position of red clump giant stars in the color
magnitude diagram, as red clump giant stars serve as a good standard
candle.

Once the images with artificial events were created, we processed them with the 
same analysis pipeline and selection criteria used in the analysis
of the actual data. We evaluated our detection efficiency as a function
of $t_{\rm E}$ in each field by simulating 20 million artificial events. The
results are shown in Fig. \ref{fig:eff}. While the detection 
efficiencies drop sharply for decreasing $t_{\rm E}$ values, the detection
efficiency is still significant at $t_{\rm E} \sim 1\,$day,
thanks to the high cadence of our survey.
Fig. \ref{fig:tE_noEff} shows the $t_{\rm E}$ distribution with the data
corrected for the detection efficiencies. This is similar to Fig. 2 of the
main paper, except that in Fig. 2, it is the predicted model 
$t_{\rm E}$ distributions that have been corrected for detection efficiencies.
The flattening of the
$t_{\rm E}$ distribution at $t_{\rm E} \simlt 2\,$days indicates a rising mass
function very low (planetary) masses, because the intrinsic
microlensing probability scales with the lens
mass $M$ as $\sim \sqrt{M}$.

\section{Tests for Systematic Biases}
\label{sec:systematic}
\hspace{1.5em}
When light curve coverage is poor or the photometry errors are large,
the light curve models can become degenerate. A model with a short
timescale (small $t_{\rm E}$) and a large $u_0$ value can look quite similar
to a longer event with a smaller $u_0$ value and a fainter source flux,
$F_{\rm s}$. The degeneracy tends to be worse for faint source stars
blended with brighter stars, and it can lead to a bias in the inferred 
event parameters.
This bias has been well studied \cite{smith2007}.  To avoid such 
systematic biases in the observed $t_{\rm E}$ distribution, we have rejected
microlensing candidates with fit parameters that are not well constrained as
described
in Section \ref{sec:analysis}. Note that the errors in $t_{\rm E}$ listed 
in Table $1$ in the main paper include the correlation with other parameters.
In order to check for possible systematic parameter biases and significant 
contamination of our event sample, we have conducted several consistency 
checks with the observed and simulated events.

The distribution of the impact parameter, $u_0$ which is purely geometric,
can be used as a test that the sample is dominated by real
microlensing events. Fig. \ref{fig:Gamma_u0} compares the
observed and simulated $u_0$ distributions for events in
different $t_{\rm E}$ ranges. We performed the Kolmogorov-Smirnov test,
which yields probabilities of 
77\%, 76\%, 51\%, 97\% and 74\% for the ranges 
$t_{\rm E}=$0.3-2.0, 2.0-13, 13-69, 89-200 days and the full sample, 
respectively. All the sub-samples are consistent with the 
simulated distributions, so there is no evidence of 
significant contamination by intrinsic variable stars 
in the $u_0$ distributions.

We also compared the input (true) $t_{\rm E,in}$ values input
into our event simulations to the $t_{\rm E,out}$ values output by
the light curve analysis. The results are shown in Fig. \ref{fig:InOut_tE}.
This test shows that the $t_{\rm E,out}$ values are a reasonably accurate
representation of the input $t_{\rm E,in}$ values, but that there is
a small systematic offset that makes the mean $t_{\rm E,out}$
systematically larger than the input (true) $t_{\rm E,in}$ values,
independent of the value of $t_{\rm E,in}$.
This implies that the bias in our measured $t_{\rm E}$ is too small
to influence our conclusions. Furthermore, the sign of the small bias
that we do see implies that we might be slightly underestimating
the number of short $t_{\rm E}$ events.

\section{Likelihood Analysis of the Substellar and Stellar Mass Functions}
\label{sec:Likelihood}
\hspace{1.5em}
The Einstein radius crossing time $t_{\rm E}$ is the only observable in the regular 
single lens microlensing event and is given by Equation (\ref{eq:lens_eq}).
Although the physical parameters of the lens and source are degenerate,
a model $t_{\rm E}$ distribution, $\Phi(t_{\rm E})$, can be calculated using 
a Monte Carlo simulation for an assumed mass function with a standard 
galactic mass density and velocity model \cite{HanGould03}.
The same method have been applied in the previous studies \cite{alc00a,sumi03}
of the stellar mass function in the Galactic bulge.

We used a Bayesian likelihood analysis to determine the 
mass function model parameters for both the power-law and 
log-normal mass functions. For the stellar and brown dwarf
mass functions this was done both with and without the 
events with $t_{\rm E}\le2\,$day events, and we found that 
the differences in the implied stellar and brown dwarf
mass functions was negligible, as shown in Table \ref{tbl:MF}.
Fig. \ref{fig:likelihood_alpha3_noPL} shows the
likelihood distribution for the mass function slope in the
brown dwarf regime, $\alpha_3$, for the power-law model.
This analysis finds $\alpha_3=-0.48_{-0.29}^{+0.37}$,  
which is slightly higher than, but consistent with, previous estimates of 
$\alpha_3=0$-$0.3$ \cite{Cruz2007,Metchev2008,Reyle2010}.
Note that this $\alpha_3$ values does not change when we 
consider additional mass function models in the following section. 

Fig. \ref{fig:likelihood_fitChabrier} is a contour plot of the
likelihood distribution for the log-normal mass function parameters
$M_{\rm c}$ and $\sigma_{\rm c}$. The contours indicate the 68\% 
and 95\% confidence intervals and the cross indicates the maximum
of the two-dimensional likelihood distribution. The log-normal
parameter values determined by this analysis are a
mean mass $M_{\rm c}=0.12\pm0.03$ and a log-normal width of
$\sigma_{\rm c}=0.76_{-0.16}^{+0.27}$. These are consistent with the
values of $M_{\rm c} = 0.079_{-0.016}^{+0.021}$ and 
$\sigma_{\rm c} = 0.69_{-0.01}^{+0.05}$ determined in the previous work \cite{Chabrier2003}.

The power-law and log-normal models indicate 
that the number of brown dwarfs is 
$0.73_{-0.19}^{+0.22}$ and $0.70_{-0.30}^{+0.19}$ 
(respectively) times the number of main sequence stars in the mass range
$0.08\le M/M_\sun \le 1.0$, which we denote by $N_*$. 
These values are higher than the value of 0.2 for the model (3)
in Section \ref{sec:Discontinued}.
The best model 
$t_{\rm E}$ distributions, $\Phi(t_{\rm E})$, are indicated as dashed lines 
in red and blue for the power-law and log-normal models, respectively, 
in Fig. 2 of the main paper and Fig. \ref{fig:tE_noEff}.  As one 
can see, the both models represent data very well for $t_{\rm E}\ge 2$,
but there is a significant excess at $t_{\rm E} \le 2\,$days. For $t_{\rm E} \le 2\,$days,
we have detected 10 events compared to the predictions of 1.5 and 2.5 events without the
planetary mass population for the power-law and log-normal models, respectively. 
The Poisson probability of detecting 10 events when only 1.5 or 2.5  events 
are predicted are $4\times10^{-6}$ and $3\times10^{-4}$ for
the power-law and log-normal models, respectively.
At $t_{\rm E} \le 1\,$days, we observe 3 events compared to predictions
of 0.12 and 0.24 events for the planet-free power-law and log-normal models,
respectively. The contributions of the best fit planetary mass distributions
are indicated by the red and blue dotted curves for the power-law and
log-normal mass function respectively, and the full distributions
including lens objects of all masses are given by the red and blue solid
curves. Both models provide an excellent fit to the observed  $t_{\rm E}$
distribution.

\section{Likelihood Analysis with the Other Mass Functions}
\label{sec:Likelihood2}
\hspace{1.5em}

In this section we present the likelihood analysis for mass functions
that differ from the models (1) and (2) shown in the main paper. 
Section~\ref{sec:Discontinued} presents the results for an alternative
stellar and brown dwarf mass function that is discontinuous near the
brown dwarf-stellar boundary, and in Section~\ref{sec:PLPower}
we consider a planetary power-law mass function. This is likely to
be more realistic than the $\delta$-function model, but its parameters
are difficult to measure with the current data set.

\subsection{Discontinuous Substellar Mass Function}
\label{sec:Discontinued}
\hspace{1.5em}
In the main paper, we used two models, (1) the power-law mass function (MF) measured
in the galactic bulge \cite{Zoccali2000} with a continuous brown dwarf 
MF that was fit to the observed $t_{\rm E}$ distribution, and (2) a log-normal MF 
fitted to the data, with best fit parameters   are consistent 
with the values found with previous work \cite{Chabrier2003}.

Here, we present the likelihood analysis with model (3), a discontinuous MF
\cite{Thies2007,Thies2008} which uses
the universal Galactic IMF, $\psi$, based on the average of various stellar
clusters \citep{Kroupa2001} with
$\alpha_1=2.3$ for $0.5M_\sun <M$,
$\alpha_2=1.3$ for $0.075M_\sun <M<0.5M_\sun$, and 
$\alpha_3=0.3$ for $0.01M_\sun  <M<0.075M_\sun$. But this MF is
discontinuous at the hydrogen burning limit $M_{\rm HBL}=0.075M_\sun$,
with a discontinuity of a factor of 
$R_{\rm HBL}=\psi_{\alpha_3}(M_{\rm HBL})/\psi_{\alpha_2}(M_{\rm HBL}) =$0.2-0.3.
We assume that stars heavier than $1M_\sun$ have evolved to become stellar
remnants, as we have done with the other mass functions. The steeper slope,
$\alpha_1=2.3$, implies fewer massive stellar remnants.
In the planetary regime, we use a $\delta$-function MF just as with models (1) and (2).
Fig. \ref{fig:likelihood_Kroupa} shows the result of likelihood 
analysis for $R_{\rm HBL}=0.3$. The resulting parameters are
$M/\msun  = 1.9_{-0.9}^{+1.4} \times 10^{-3}$ and
$\Phi = 0.50_{-0.10}^{+0.11}$ as shown in Table \ref{tbl:MF}. 
In this model, planetary mass objects with mass of $1.9_{-0.9}^{+1.4}$ 
Jupiter mass are $1.3_{-0.4}^{+0.7}$ times as frequent as main sequence stars.
With a slightly stronger mass function discontinuity, $R_{\rm HBL}=0.2$, we
find  similar values:
$M/\msun   = 2.2_{-1.0}^{+1.6} \times 10^{-3}$ and
$\Phi = 0.51_{-0.10}^{+0.10}$.
Fig. \ref{fig:tE_Kroupa} presents the best fit  $t_{\rm E}$ distribution 
for the case of $R_{\rm HBL}=0.3$. One can see that this model generates a 
narrower $t_{\rm E}$ distribution. It has fewer long ($\ge$40 day) and short 
($\le$10 day) events compared to the peak ($\sim$20 days), than 
models (1) and (2) shown in the main paper. This is because this model
has a steeper power law for high masses, $0.5M_\sun <M$, and 
fewer brown dwarfs. This model provides a worse fit to the data than
models (1) and (2), with a maximum likelihood value that is a factor of
25 worse than model (1). However,the discontinuous MF model (3)
does have one fewer model parameter that is fit to the data and it employs
a mass function at the high end, which provided a worse fit to long timescale
events in previous studies\cite{sumi03,bennett_bh}. Since the relatively poor fit at
the high mass end contributes to the poor likelihood value, we consider this
model of the mass function of low-mass objects to be acceptable.
The main conclusion, however, is that this mass
function does not alter our conclusion that a new population of 
planetary mass objects is needed to explain the data, although it does
imply a slightly higher mass for these objects.

\subsection{Power-law Planetary Mass Function}
\label{sec:PLPower}
\hspace{1.5em}
In the main paper, we used a $\delta$-function mass function for
the planetary mass objects, but this is not likely to be realistic. 
Exoplanets are known to span a wide range of masses, and lower mass
planets are much more easily ejected from their host stars in 2-body 
encounters. So, we have also considered a power-law form for the planetary
part of the mass function, with an index of $\alpha_{\rm PL}$ for the mass 
range $10^{-5}\le M/M_\sun \le 0.01$. This is assumed to be continuous
with the power-law stellar and substellar mass 
function of MF model (1) in the main paper, at $M=0.01 M_\sun$.
Our result does not depend on the lower mass limit $10^{-5}M_\sun$, 
because our sensitivity to events with $M < 10^{-4} M_\sun$ is quite
small, due to the very low the detection efficiency at small $t_{\rm E}$.
The power-law index, $\alpha_3$, for the brown dwarf mass regime,
$0.01 \le M/M_\sun \le 0.08$, was fit with $\alpha_{\rm PL}$, simultaneously.

Fig. \ref{fig:likelihood_PLPow} shows the likelihood function for 
$\alpha_3$ and $\alpha_{\rm PL}$, and the resulting best fit parameters are
$\alpha_3 = 0.49_{-0.27}^{+0.24}$ and
$\alpha_{\rm PL} = 1.3_{-0.4}^{+0.3}$.
This $\alpha_3$ value is consistent with our result for MF model (1)
with and without $\delta$-function planetary mass function
(see Table  \ref{tbl:MF}).
The value $\alpha_{\rm PL}$, which is as steep as $\alpha_2$, is much steeper than $\alpha_3$,
indicating that this planetary mass objects are separate population from
the brown dwarfs. This model also predicts a larger, but poorly constrained,
number of planetary mass objects per star with $N_{\rm PL}/N_* = 5.5{+18.1\atop -4.3}$, 
due to our poor sensitivity to very low-mass lenses.

Fig. \ref{fig:tE_PLPow} shows the best fit $t_{\rm E}$ distribution compared with
data and the best fit model (1) in the main paper for a comparison. Although 
this power-law planetary mass function model has a maximum likelihood that is
a factor of 0.75 smaller than model (1), it has one fewer model parameter, so
formally, it is a slightly better fit. This implies that the current data
set is not able to constrain the shape of the mass function at sub-Jupiter
masses. 

\section{Relative Frequencies of Unbound/Distant and Bound Planets}
\label{sec:Frequency}
\hspace{1.5em}
Our sample includes three previously known events in which
a planet and its host star are both detected: MOA-2007-BLG-192
\citep{bennett08,kubas2010}, OGLE-2007-BLG-368 \citep{Sumi2010}, and 
OGLE-2007-BLG-349 \citep{Gould2010}. 

The short timescale events caused by planets bound to host stars
with a separation of $<10-20$ AU are detected as binary lens systems
with a very small mass ratio. The events in our sample are detected
as single lens events, because the planetary mass lenses are sufficiently
isolated from any host star that the microlensing effect of the host star
cannot be seen. These isolated planetary mass lenses can be either
unbound or in distant orbits about a host star, and they can be distinguished
from planetary mass lenses detected in binary events 
in two possible ways. The binary lens systems
reveal the presence of a nearby host star by showing the
effect of a planetary caustic in their light curves or by showing a
low-magnification bump due to lensing by the host 
star \citep{Han2003,Han2005,Han2009}. 
We searched for the signature of host stars in our short events by fitting
binary light curves to the data and found that none of 
the 10 selected events 
with $t_{\rm E}<2$ days shows any significant evidence of host stars (see bellow
for their limits). However, our analysis has revealed 
three short events, which fail our selection criteria, that have clear
binary lens caustic crossing features, as well as a very low
amplitude signal due to lensing
by the host star. Detailed modeling of these 3 events indicates wide binary
models with lens mass ratios of $q < 0.05$ for two of these events and
a close binary for the third (Bennett et al. in preparation). 

Thus, the total number of events in this sample in which both a planet 
and its host star are detected is 5. However, for two of these events, 
MOA-2007-BLG-192 and OGLE-2007-BLG-349, the planet is detected via light 
curve features due to the central caustic. This means that these planets have 
been detected only because of the gravitational effect of the host stars. 
For the other 3 planet plus host star events, the planet would have been 
detected without the influence of the host stars. Of course, there are 
likely to be some planets even closer to their host stars that are 
undetectable (by microlensing) 
because of the presence of the host star, so it is reasonable to 
presume that there are 5 planetary mass objects in binary lens systems in 
this sample that were detected with approximately the same efficiency 
as the planetary mass objects detected as isolated lens systems.
This allows us to estimate the ratio of the new population of unbound or distant
planetary mass objects to the planets found with host stars. The power-law mass function
implies 11 isolated planetary mass lenses (that comprise our unbound or
distant planetary mass sample) in our full event sample, so this ratio
is $11/5 = 2.2$. With the log-normal mass function, the number of events
due to planetary mass objects is $\sim 11$, but 3 of these are from the
low-mass tail of the brown dwarf mass function. This leaves 8 unbound or distant
planetary mass objects for a ratio to the planets with detectable host stars of $8/5 = 1.6$.

The frequency of planets per star was measured by 
microlensing\citep{Gould2010} for the planet-to-star
mass-ratio interval $-4.5 < \log q < -2$ and separation $a\sim$2-8\,AU, to be
$d^2N_{\rm pl}(d\log q d\log s)^{-1}=0.36\pm 0.15$.
This indicates $N_{\rm pl}\sim$0.9 for this $q$ range and interval $a=1$-10\,AU.
Radial velocity (RV) surveys have found that 11.3\% of stars have 
planets with $M \sin i > 10 M_\oplus$ with periods less than 50 days \cite{Howard2010}.
The RV planets have periods that go down to about 3 days 
before the numbers really drop. So, we consider the RV survey to 
cover the period range $3 < P < 50$ day.
If we assume a typical host star mass of 0.5 solar masses, then periods of
$P=3$ and $50$ days imply semi-major axes of
$a =0.032$ and $0.21$\,AU, respectively, corresponding a range of $0.81$ in $\log(a)$.
Assuming a uniform planet distribution in $\log(a)$, we have a 0.21 bound 
planets per star between $a=0.21$ AU and $1$ AU where the microlensing numbers take over,
corresponding a range of $0.68$ in $\log(a)$.
If we assume a planet distribution in period of
${\rm d}N_{\rm pl}({\rm d}\log P)^{-1} \propto P^{0.26}$ as
measured by RV for G dwarfs \citep{cumming08}, we have a 0.29 bound planets per star.
Thus, we have a total of 1.13-1.19 bound planets per star in the
semi-major axis range $a=0.03$-$10$\,AU. This implies
1.5-1.6 times as many planetary mass objects that serve as isolated lens systems to
planets known to orbit stars. This result is nearly identical to the
microlensing-only data based argument above. 

By necessity, we define our sample of isolated planetary mass lens
objects as ones who don't
have a host star that can be detected by microlensing, but this does not
necessarily imply that host stars do not exist, which is why we interpret them
as unbound or distant planets. The lower limit on the possible
separation of a host star depends primarily on the magnification of the
observed short timescale event. These minimum host star separations,
$d_{\rm min}$, are given in terms of the Einstein radius of the hypothetical
host star, and the values for each of our $t_{\rm E} \le 2\,$day events are
listed in the last column of Table $1$ in the main paper. These are limits on the
projected separation, and the typical semi-major axis corresponding to
a host star with a separation of $d_{\rm min}$ of one of our detected
unbound or distant planetary mass object is $\sim (3\,{\rm AU})d_{\rm min}$, and therefore the
typical lower limit on the separation of a host star is $\sim 14\,$AU, with
a range of $\sim 7$--$45\,$AU depending on the event.

Our data do not indicate what fraction of this population of unbound or distant
planetary mass objects actually have host stars outside the region of detectability, but
we can use the limits from direct detection searches to estimate this 
fraction.
The Gemini Planet Imager\cite{Lafreniere2007} has set upper limits on
the number of stars with Jupiter-mass planets at semi-major axes of 10-500 AU,
with the tightest limits of $<30$\% of stars with Jupiter-mass planets 
with semi-major axes in the range 50-250 AU. Based on these results, we 
estimated that the fraction of stars with a 1 Jupiter-mass
planet with semi-major axes in the range $10\,{\rm AU} < a < 500\,{\rm AU}$
is less than 40\% assuming a uniform distribution in $\log a$.
Since the new unbound or distant 
planetary mass population is comprised of $\sim 1.8$ times
as many Jupiter-mass planets as stars, this comparison with the
direct detection limits suggests that at least 75\% of these unbound or distant
planetary mass objects are not bound to any host star.

Note that our likelihood analysis of the mass function in the planetary mass
range, shown in Fig. \ref{fig:likelihood_Kroupa}, givens a 1-$\sigma$ lower 
bound on the $\delta$-function mass of the planetary mass objects of 
approximately a Saturn mass.  
The direct imaging results don't rule out Saturn mass planets.
So, if the majority of the lens objects in the newly discovery planetary
mass population have a mass of order that of Saturn or less, then
the direct detection limits do not apply, and the majority of this new
sample can be planets bound to host stars in wide orbits with
semi-major axes of $a \simgt 7$--$45\,$AU depending on the event.

\clearpage
\setcounter{figure}{0}
\setcounter{table}{0}
%-------------FIG.S1--------------------
\begin{figure*}
%\epsscale{0.5}
\includegraphics[angle=-90,scale=0.33,keepaspectratio]{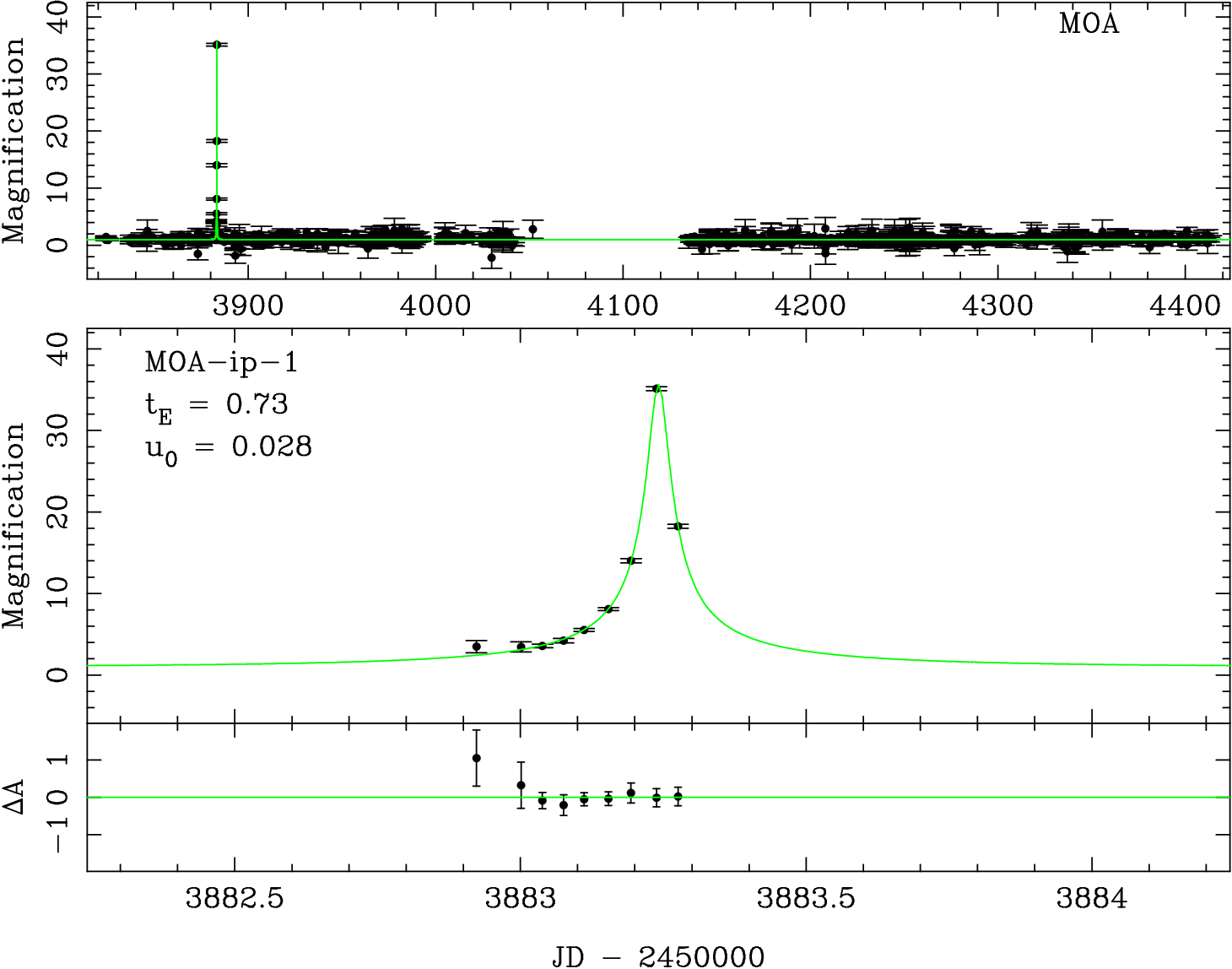}
\includegraphics[angle=-90,scale=0.33,keepaspectratio]{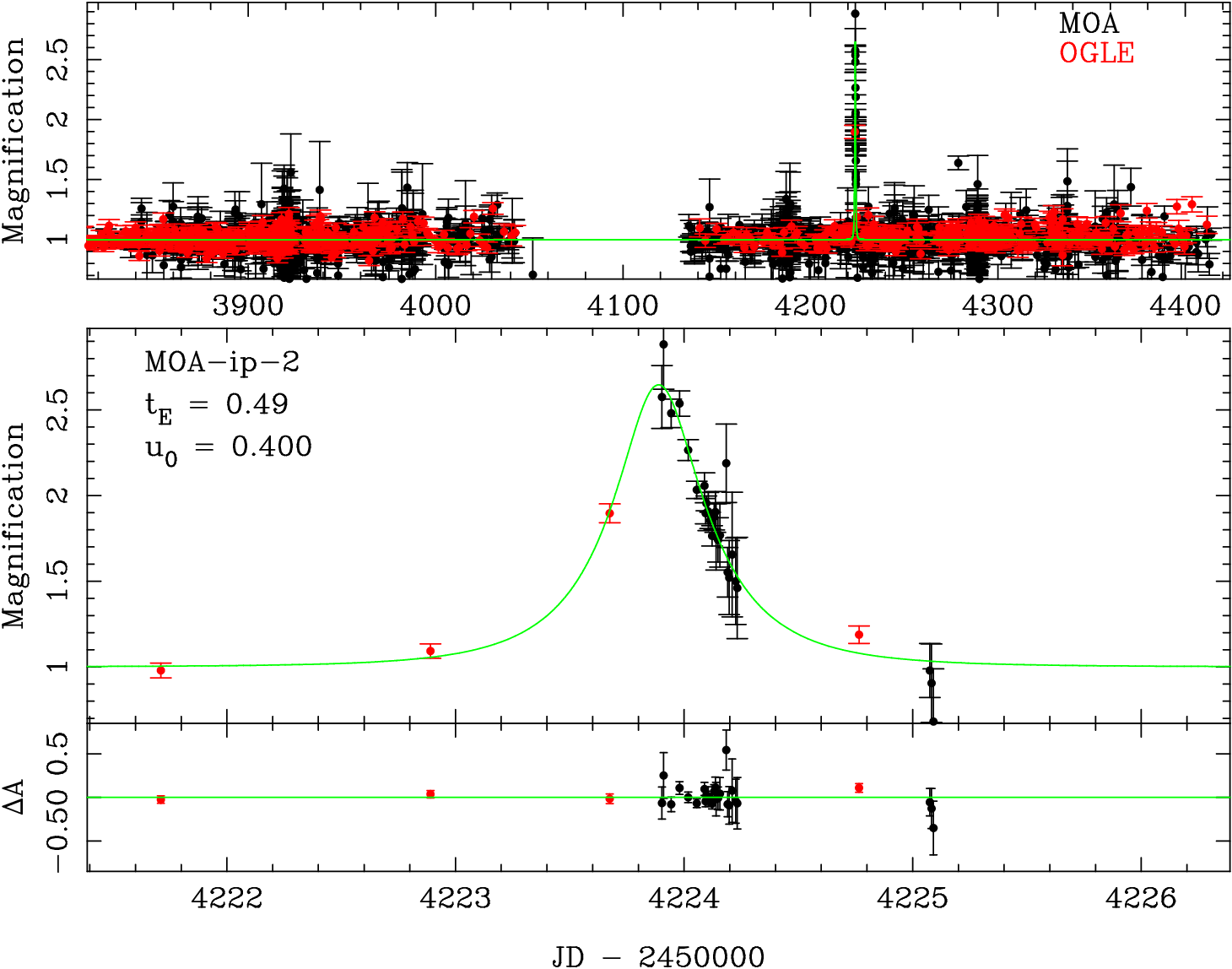}
\includegraphics[angle=-90,scale=0.33,keepaspectratio]{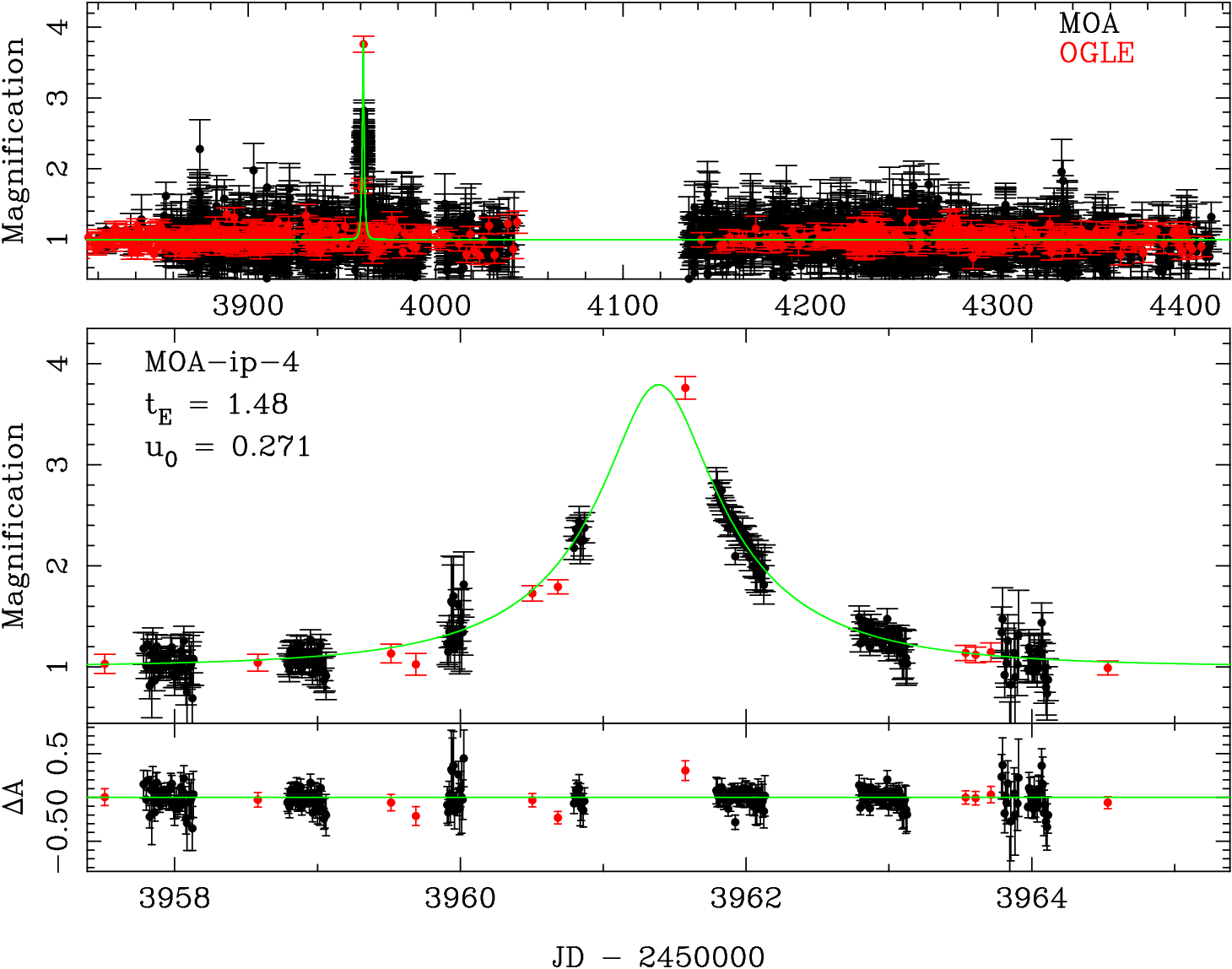}
\includegraphics[angle=-90,scale=0.33,keepaspectratio]{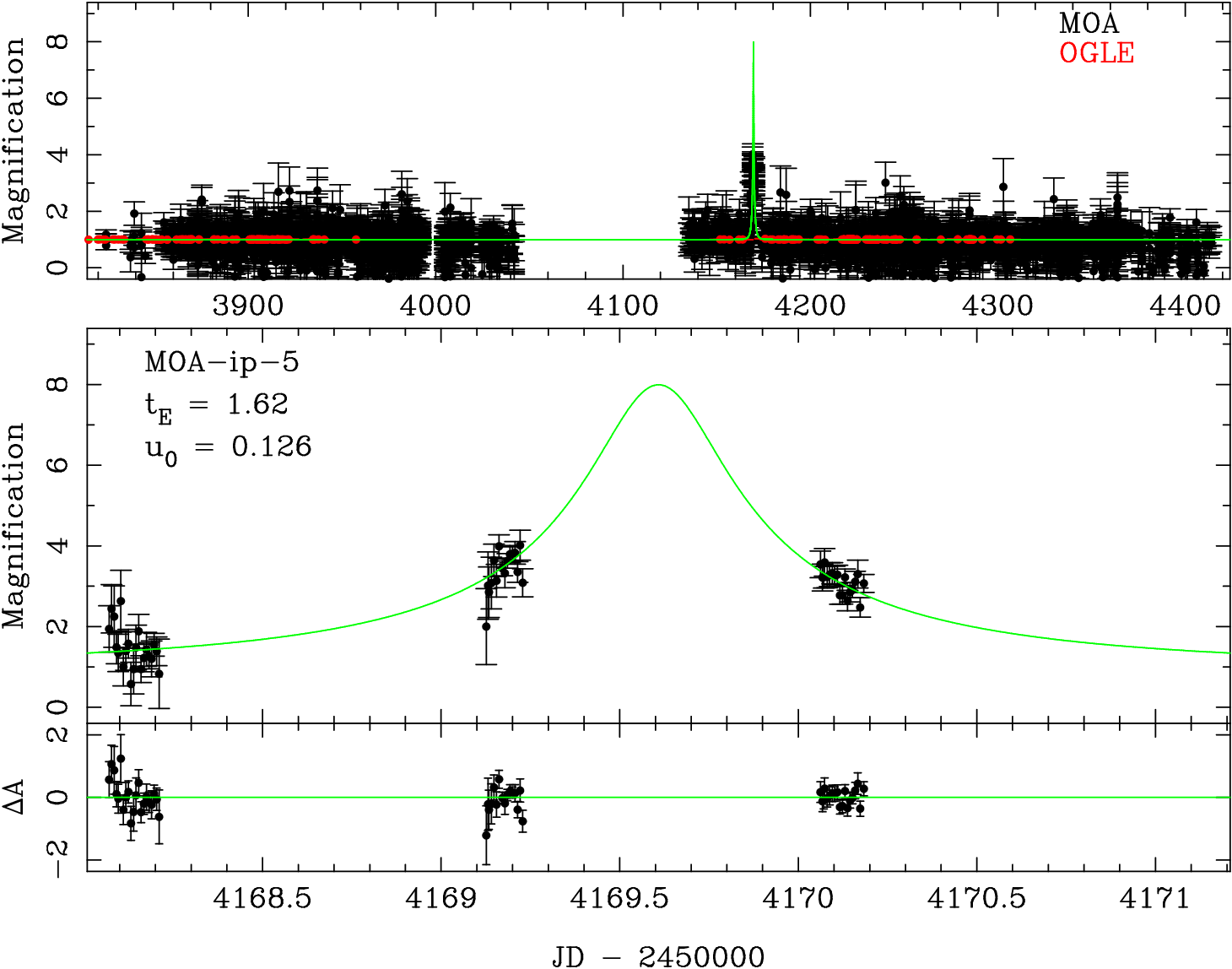}
\caption{
Light curves of the 8 short microlensing events with $t_{\rm E} \le 2$days that are
not shown in the main paper. MOA data are in black and OGLE data 
are in red with error bars indicating s.e.m. For each event, the upper panel 
shows the full two-year light curve, and the middle and lower panels show
a zoom-in around the peak and 
the residual from the best fit model, respectively.
Although the light curve coverage of event MOA-ip-5 is somewhat sparse,
it has 69 data points during the peak ($t_0\pm t_{\rm E}$), the second largest number in our short 
event sample (see Table $1$ in the main paper) and our light curve model fits support
the conclusion that this is a microlensing event.
%Light curve of gb11-R-9-27004 over the whole two-year coverage (bottom-panel), around 
%the magnification (top panel) with the residual from the best fit model.
\label{fig:lc}}
\end{figure*}

%-------------FIG.S1--------------------
\begin{figure*}
%\epsscale{0.5}
\includegraphics[angle=-90,scale=0.33,keepaspectratio]{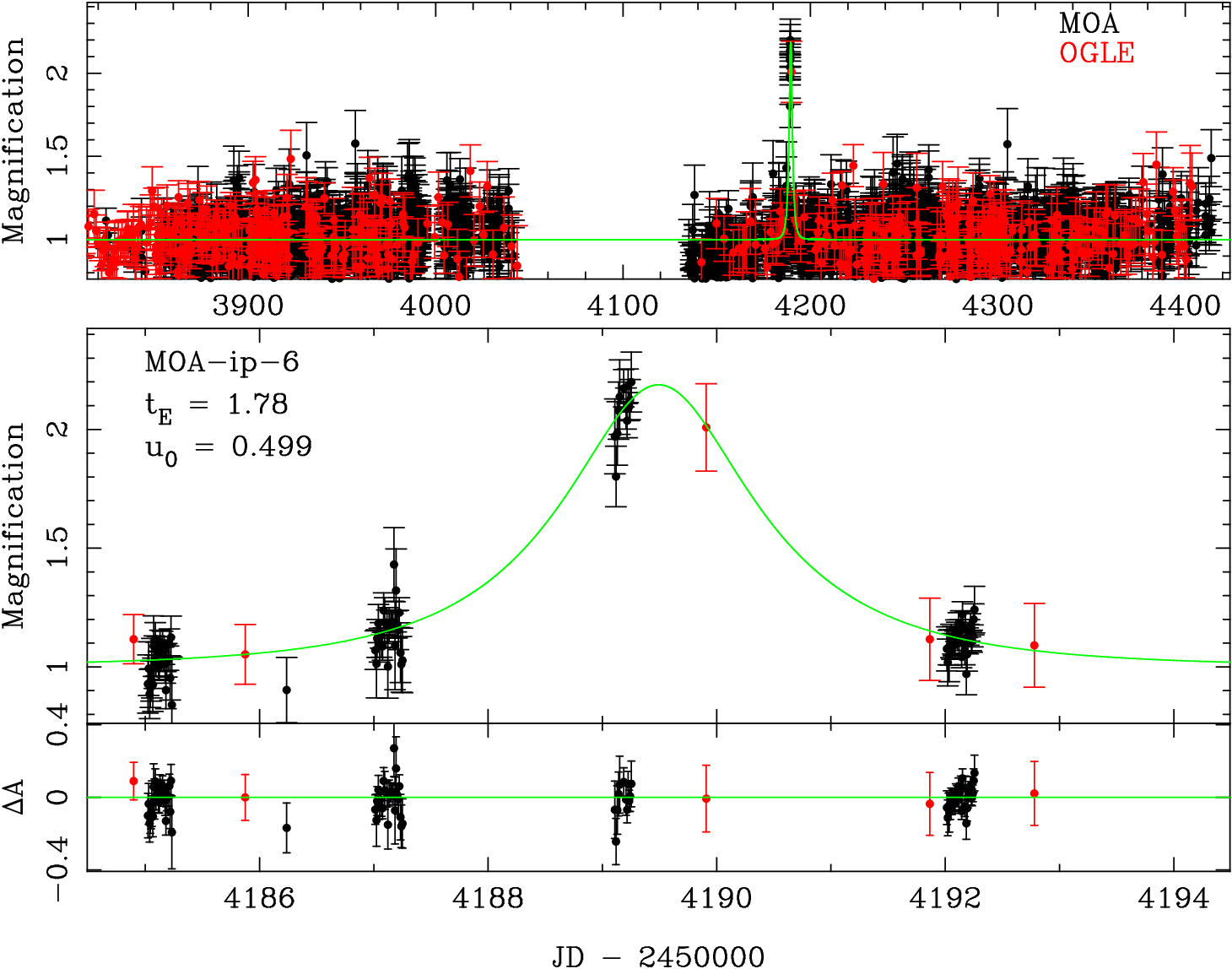}
\includegraphics[angle=-90,scale=0.33,keepaspectratio]{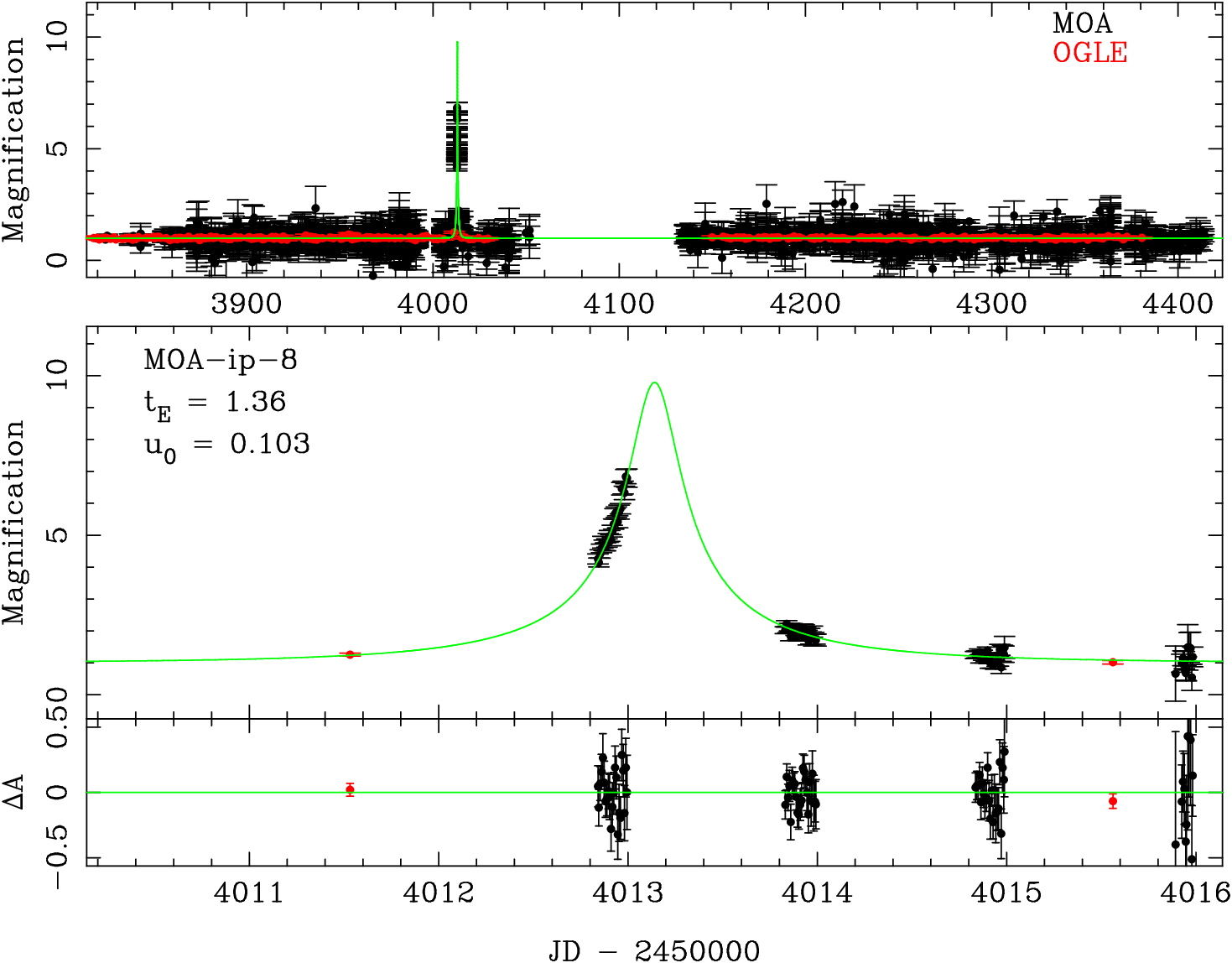}
\includegraphics[angle=-90,scale=0.33,keepaspectratio]{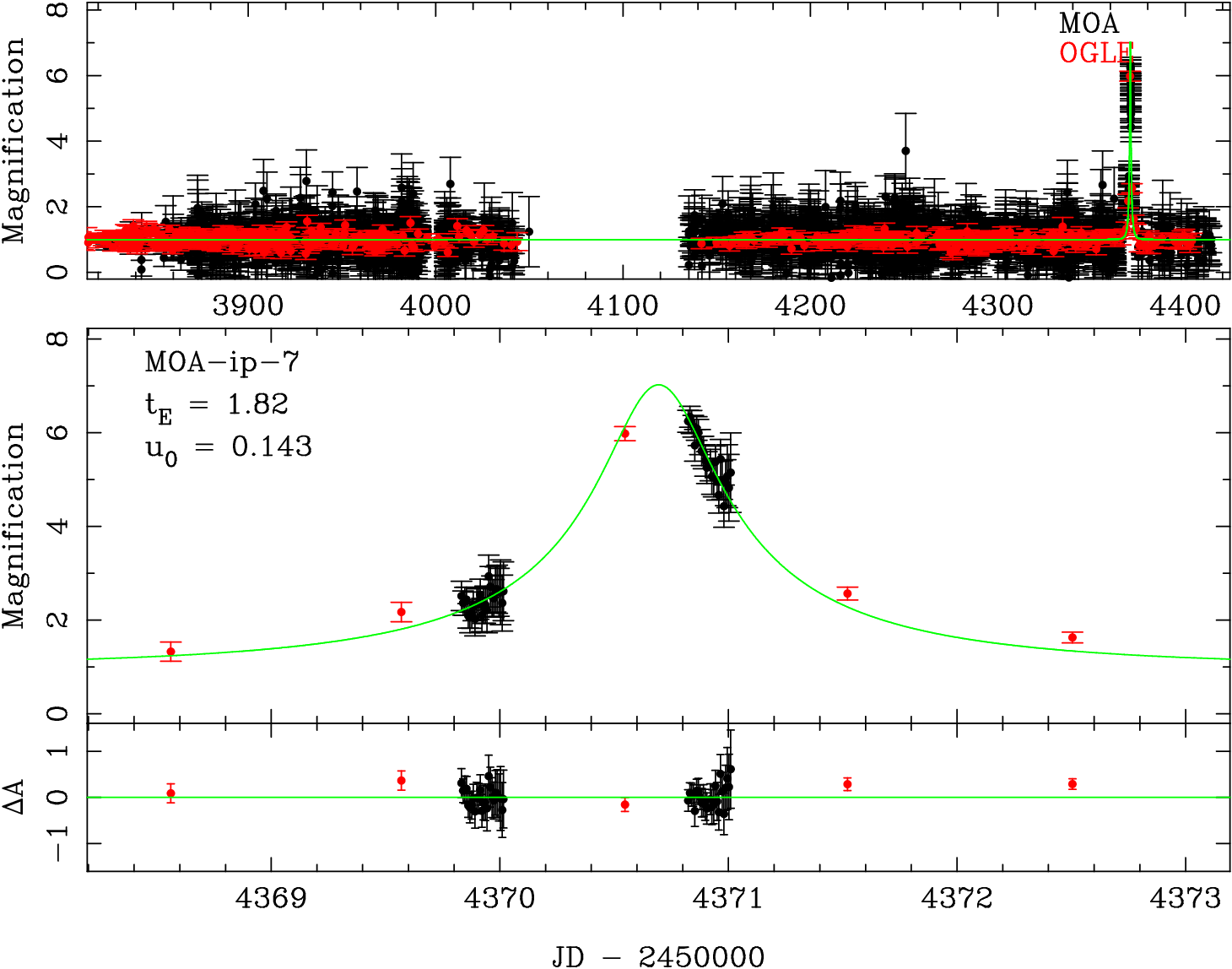}
\includegraphics[angle=-90,scale=0.33,keepaspectratio]{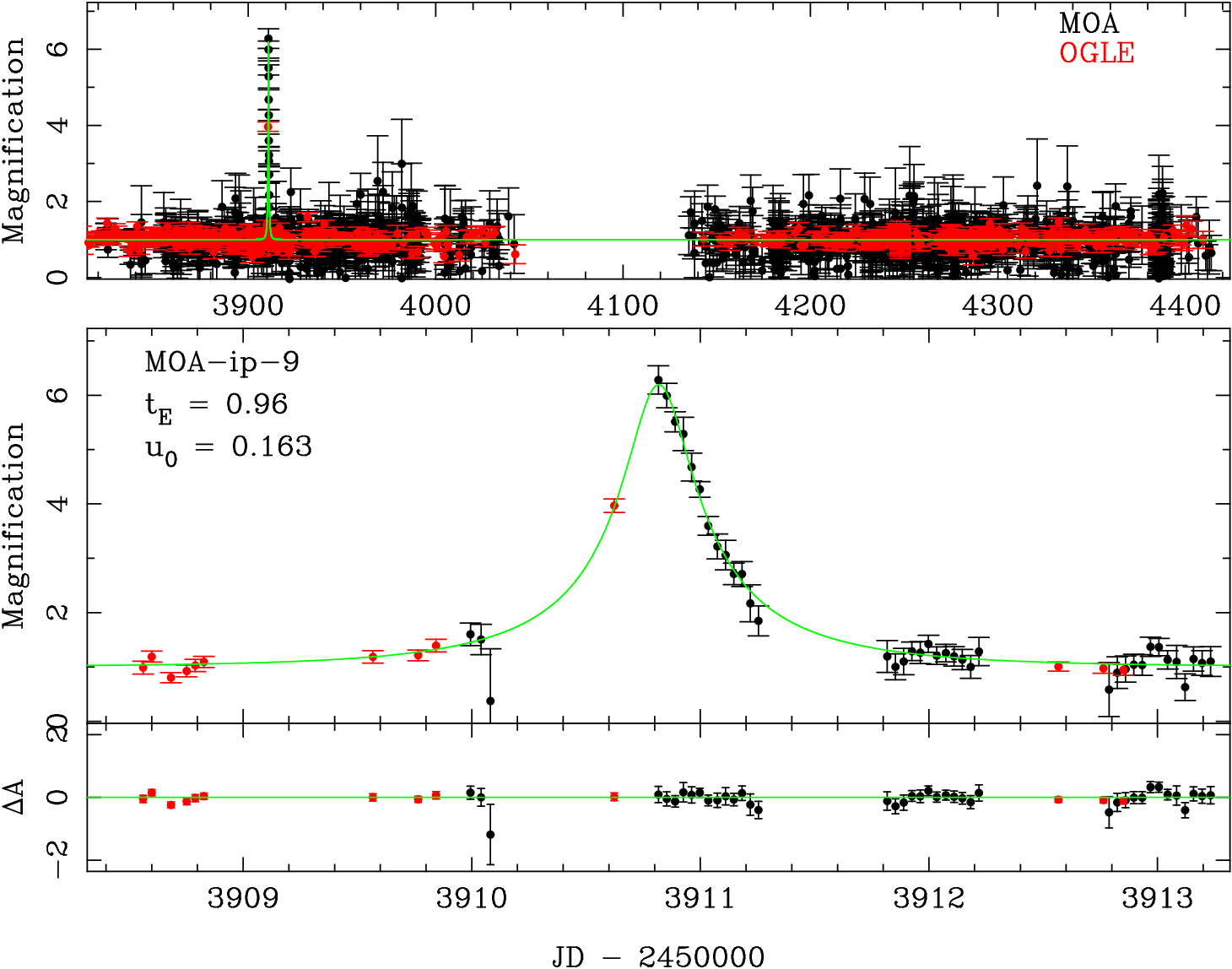}
\contcaption{continued.}
\end{figure*}

%-------------FIG.2-------------------- 
%\begin{figure}
%\epsscale{0.5}
%\includegraphics[angle=-90,scale=0.5,keepaspectratio]{plot_tE.eps}
%\caption{
%Timescale, $t_{\rm E}$, distribution of 474 microlensing events (histogram)
%and the best-fit models with Power-law (red lines) and Log-normal (blue) mass function.
%In each model, dashed lines indicate models for stellar, stellar remnant and brown 
%dwarf populations. Dotted lines represent the Planetary-mass population (PL).
%Solid lines are the sum of these populations.
%\label{fig:tE}}
%\end{figure}

%-------------FIG. S2--------------------
\begin{figure}
%\epsscale{0.5}
\includegraphics[angle=-90,scale=0.6,keepaspectratio]{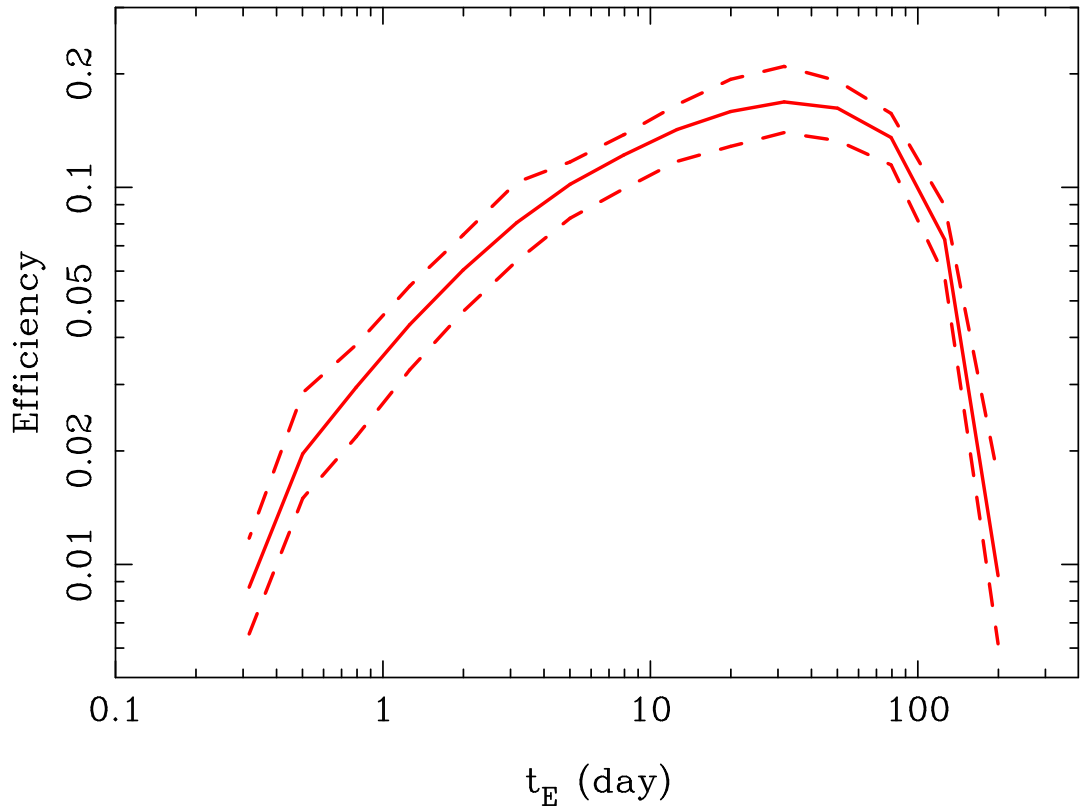}
\caption{
The detection efficiencies of our experiment as a function of $t_{\rm E}$
for the source stars down to $I=20.0$ mag. Solid and dashed lines indicate
the mean, minimum and maximum efficiencies of all fields.
\label{fig:eff}}
\end{figure}

%-------------FIG.S3-------------------- 
\begin{figure}
%\epsscale{0.5}
\includegraphics[angle=-90,scale=0.6,keepaspectratio]{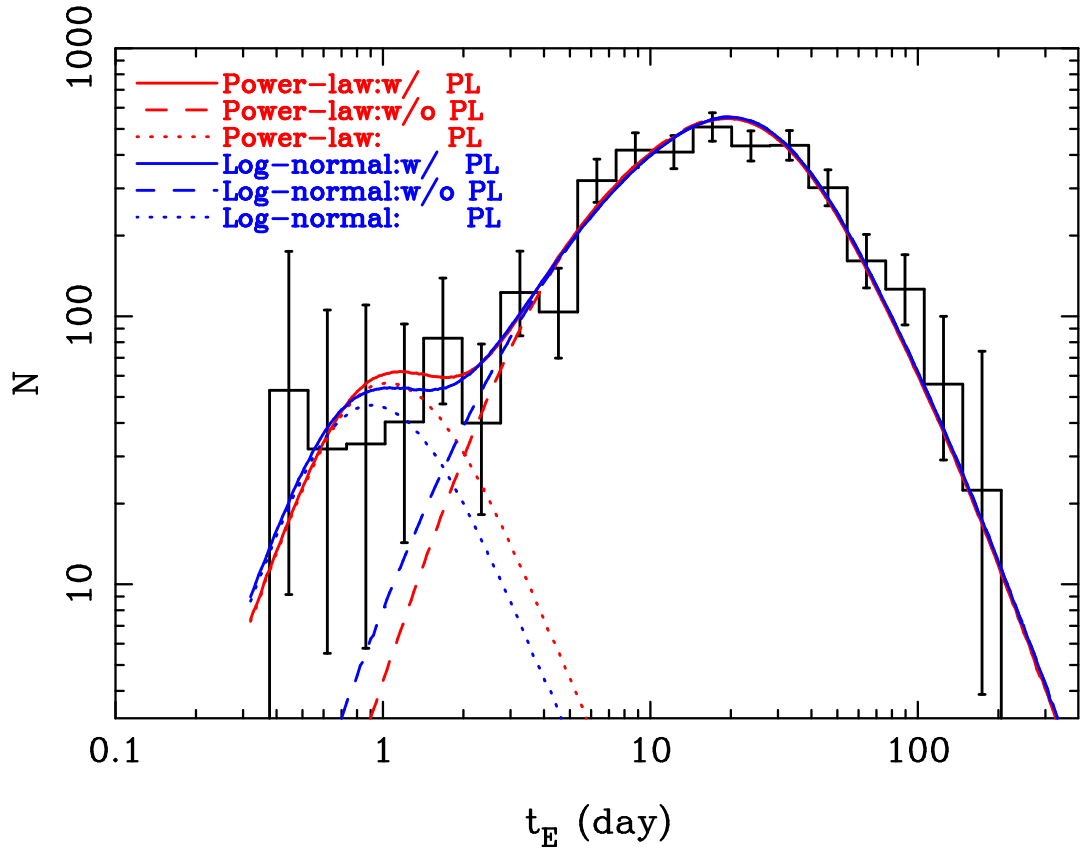}
\caption{
Observed and theoretical distributions of the event timescale $t_{\rm E}$. The black histogram
represents the observed 474 microlensing events. This is the same data that is
shown in Fig. 2 of the main paper, but here it is the data and not the models that
are corrected for detection efficiencies. The error bars indicate s.e.m..
The best-fit models with the (1) Power-law and (2) Log-normal mass functions
are indicated in red and blue, respectively.
For each model, dashed lines indicate models for stellar, stellar remnant and brown 
dwarf populations and dotted lines represent the Planetary-mass population (PL).
Solid lines are the sum of these populations.
\label{fig:tE_noEff}}
\end{figure}

%-------------FIG.3-------------------- 
%\begin{figure}
%\epsscale{0.5}
%\includegraphics[angle=-90,scale=0.5,keepaspectratio]{plot_effI.eps}
%\caption{
%SF. The detection efficiency as a function of $I$-band source magnitude
%for different $t_{\rm E}$ bins.
%\label{fig:effI}}
%\end{figure}

%-------------FIG.S4-------------------- 
\begin{figure}
%\epsscale{0.5}
\includegraphics[angle=-90,scale=0.65,keepaspectratio]{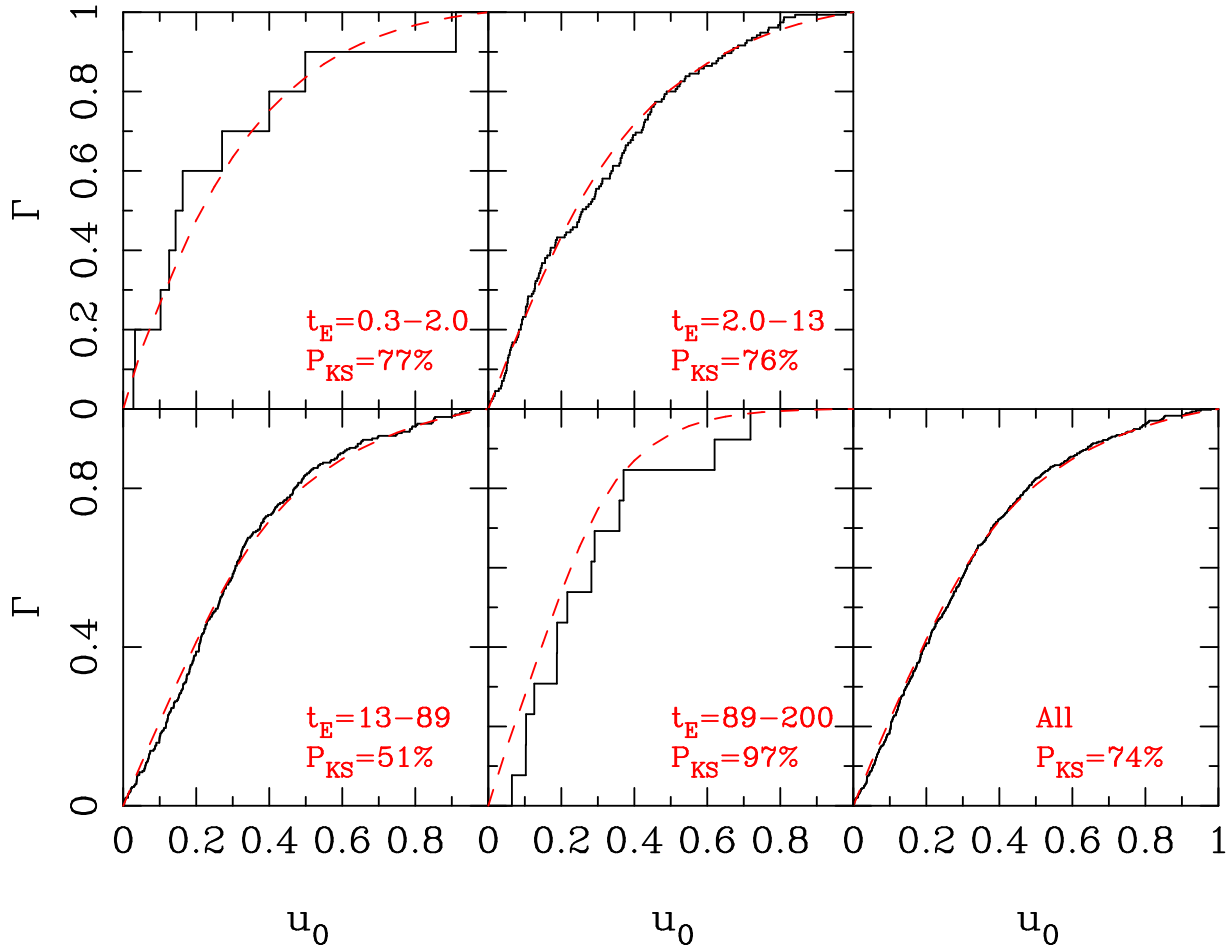}
\caption{
Cumulative distribution of the impact parameters, $u_0$, for the 
observed sample of 474 microlensing 
events (histograms) split into different $t_{\rm E}$ bins. Dashed red curves indicate 
the expected distribution from the simulation in each $t_{\rm E}$.
The probability that the observed samples are compatible with the model
are estimated with the Kolmogorov-Smirnov test, which yields 
probabilities, $P_{\rm KS}$, of
77\%, 76\%, 51\%, 97\% and 74\% for the $t_{\rm E}$ ranges 
$t_{\rm E}=$0.3-2.0, 2.0-13, 13-69, 89-200 days and all sample, respectively.
The detection efficiency is lower for events with large $u_0$ (low magnification).
For the longest $t_{\rm E}$ bin, we expect relatively 
large S/N even for large $u_0$. However, we have set a window of 
120-days in cut-1, which the estimates the S/N by comparing to the 
deviation outside of the window. For the long $t_{\rm E}$, the tails of 
events extend outside of the window and decrease the S/N. 
This decreases the detection efficiency for events large $u_0$ and $t_{\rm E}$.
\label{fig:Gamma_u0}}
\end{figure}

%-------------FIG.4--------------------
%\begin{figure}
%\epsscale{0.5}
%\includegraphics[angle=-90,scale=0.5,keepaspectratio]{plot_Gamma_I.eps}
%\caption{
%Source $I$-band magnitude distribution of 474 microlensing events (histogram) 
%for different $t_{\rm E}$ bins. Solid curves indicate the expected distribution
%from the simulation normalized to the number of events in each $t_{\rm E}$ sample.
%\label{fig:Gamma_I}}
%\end{figure}

%-------------FIG.S5--------------------
\begin{figure}
%\epsscale{0.5}
\includegraphics[angle=-90,scale=0.7,keepaspectratio]{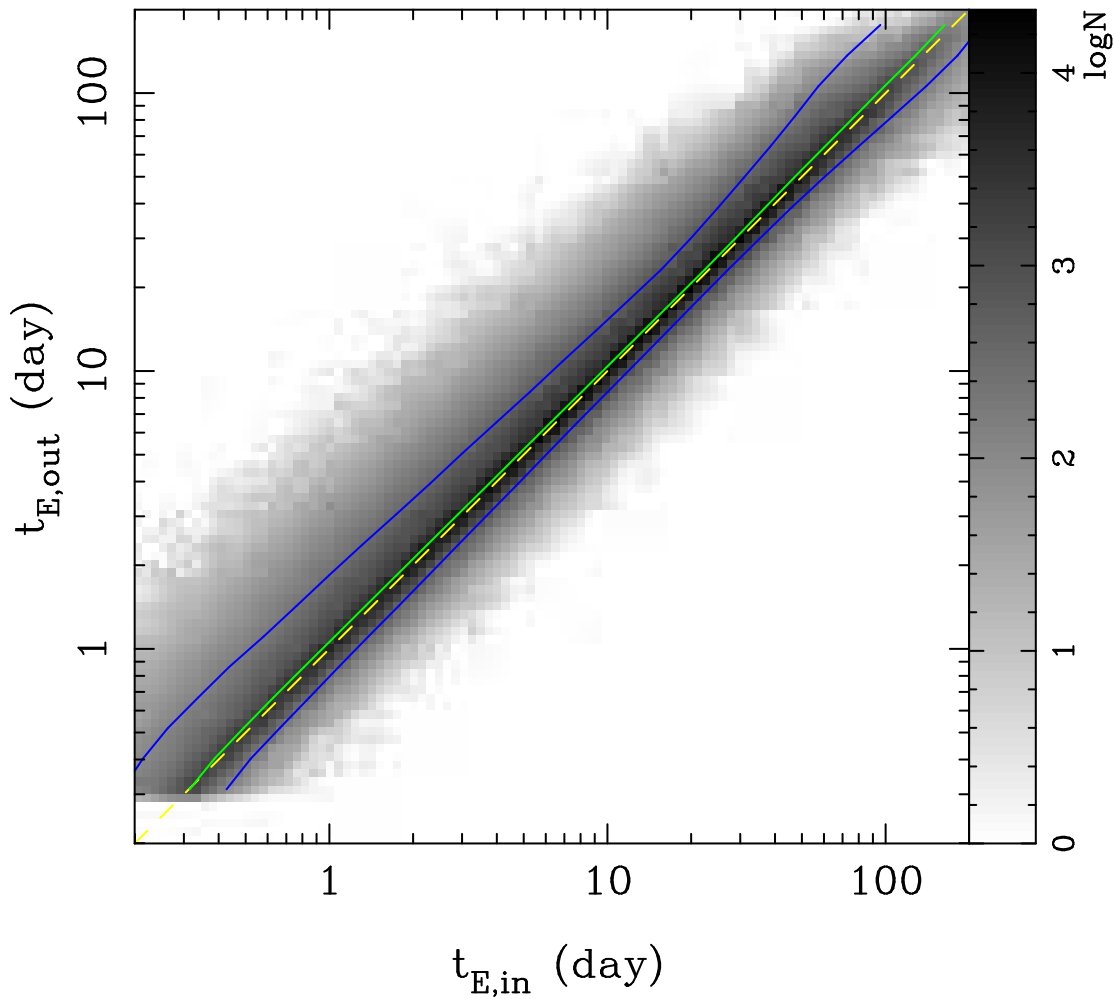}
\caption{
The output timescale, $t_{\rm E,out}$, as a function of input $t_{\rm E,in}$ of artificial 
events in our event simulations. The number density of events in each bin is 
shown as a logarithmic gray-scale, as indicated on the right.
The artificial evens were generated uniformly in magnitude and $\log(t_{\rm E})$.
Only events which passed the selection criteria are presented, and we have
not weighted by the Luminosity function or true  $t_{\rm E,in}$ distribution. 
The yellow dashed line indicates $t_{\rm E,in}=t_{\rm E,out}$, and
the green solid line represents the mean of $t_{\rm E,in}$ as a function of a mean of
$t_{\rm E,out}$. We found $t_{\rm E,in}$ is systematically $\sim5$\% 
smaller than $t_{\rm E,out}$ regardless of $t_{\rm E}$.
The blue solid lines indicate 90\% interval in each  $t_{\rm E,out}$ bin.
\label{fig:InOut_tE}}
\end{figure}

%-------------FIG.5--------------------
%\begin{figure}
%\epsscale{0.5}
%\includegraphics[angle=-90,scale=0.6,keepaspectratio]{plot_InOut.u0.eps}
%\caption{
%Same as Figure \ref{fig:InOut_tE} for the impact parameter, $u_{\rm 0}$.
%(See caption of Figure \ref{fig:InOut_tE})
%\label{fig:InOut_u0}}
%\end{figure}

%-------------FIG.6--------------------
%\begin{figure}
%\epsscale{0.5}
%\includegraphics[angle=-90,scale=0.6,keepaspectratio]{plot_InOut.I.eps}
%\caption{
%Same as Figure \ref{fig:InOut_tE} for the $I$-band source magnitude, $I$.
%(See caption of Figure \ref{fig:InOut_tE})
%\label{fig:InOut_I}}
%\end{figure}

%-------------FIG.S6--------------------
\begin{figure}
%\epsscale{0.5}
\includegraphics[angle=-90,scale=0.6,keepaspectratio]{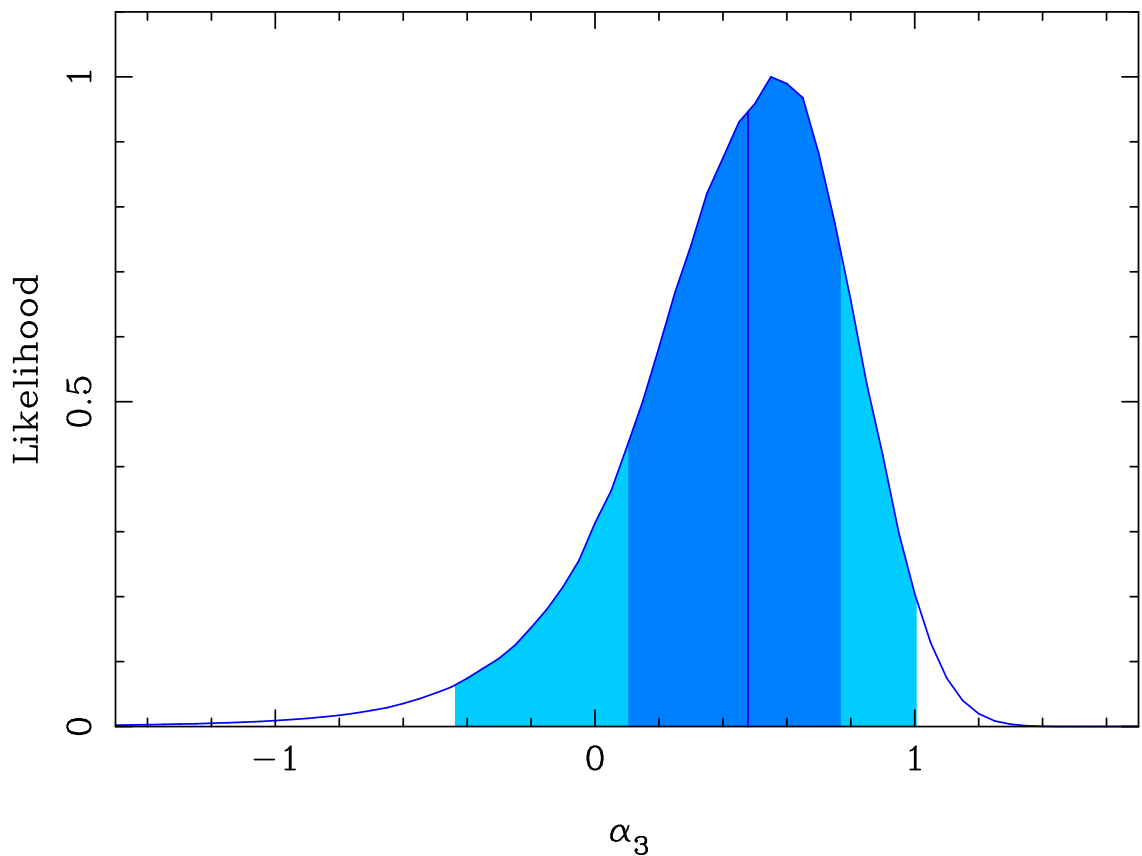}
\caption{
Likelihood distribution for the Power-law index $\alpha_3$ in brown dwarf regime without 
the planetary mass population, including only
the events with $t_{\rm E} > 2\,$days.
The vertical lines and colored regions indicate the median ($\alpha_3=0.48$) and 
68\% ($\alpha_3=0.10$ and 0.77) and 95\% ($\alpha_3=-0.44$ and 1.01) confidence intervals.
\label{fig:likelihood_alpha3_noPL}}
\end{figure}

%-------------FIG.S7--------------------
\begin{figure}
%\epsscale{0.5}
\includegraphics[angle=-90,scale=0.6,keepaspectratio]{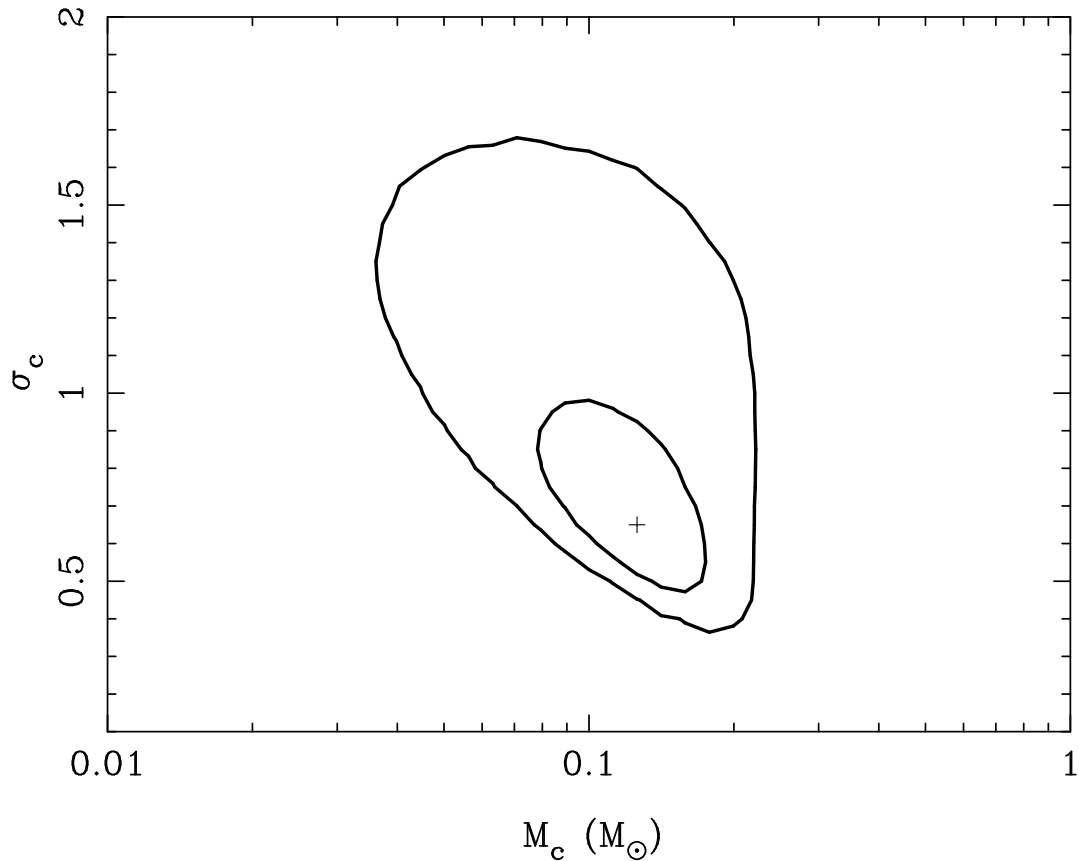}
\caption{
Likelihood contours for the mean mass, $M_{\rm c}$, and variance, $\sigma_{\rm c}$ of 
log-normal mass functions without the planetary mass population, including only
the events with $t_{\rm E} > 2\,$days.
Two contours indicate the 68\% and 95\% confidence intervals.
"+" indicate the maximum likelihood point. 
The median and 68\% confidence intervals are $M_{\rm c}=0.12_{-0.03}^{+0.03}$ and 
$\sigma_{\rm c}=0.76_{-0.16}^{+0.27}$ and these are consistent with previous 
work\cite{Chabrier2003}.
\label{fig:likelihood_fitChabrier}}
\end{figure}

%-------------FIG.S8--------------------
\begin{figure}
%\epsscale{0.5}
\includegraphics[angle=-90,scale=0.6,keepaspectratio]{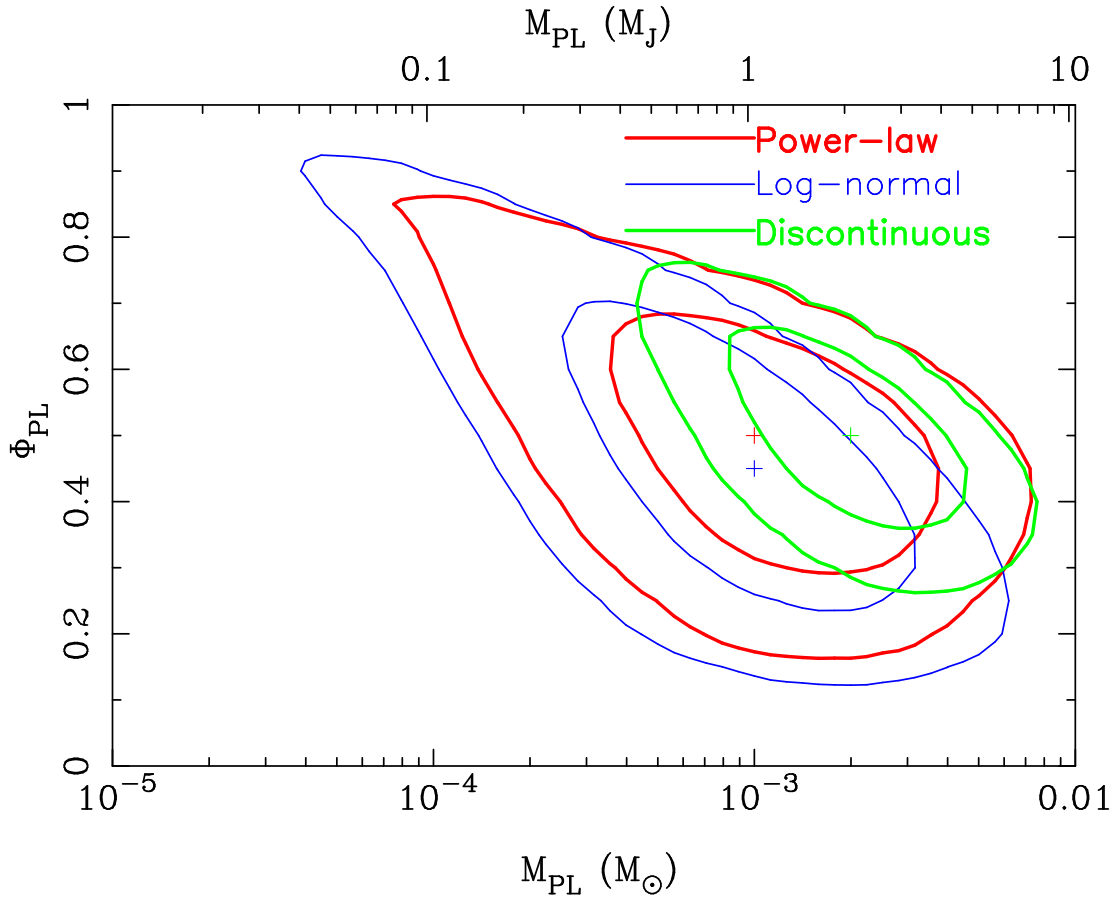}
\caption{
Likelihood contours for the planetary mass function parameters:
the fraction of all objects that are in the planetary mass population, $\Phi_{\rm PL}$,
and their mass, $M_{\rm PL}$, assuming a $\delta$-function mass function.
The two contours in each color indicate the 68\% and 95\% confidence levels.
 "+" indicate the maximum 
likelihood points.The red and blue contours represent the Power-law (1) and 
log-normal (2) mass functions also shown Fig. $3$ 
of the main paper. 
The green lines indicate the discontinuous mass function model (3)
\cite{Thies2007,Thies2008}, which uses the universal stellar mass function and 
discontinuous brown dwarf mass function with a power-law 
index of $\alpha_3=0.3$ and a scale of $R_{\rm HBL}=0.3$.
The planetary mass functions are consistent each other despite the
different brown dwarf mass functions.
\label{fig:likelihood_Kroupa}}
\end{figure}

%-------------FIG.S9-------------------- 
\begin{figure}
%\epsscale{0.5}
\includegraphics[angle=-90,scale=0.6,keepaspectratio]{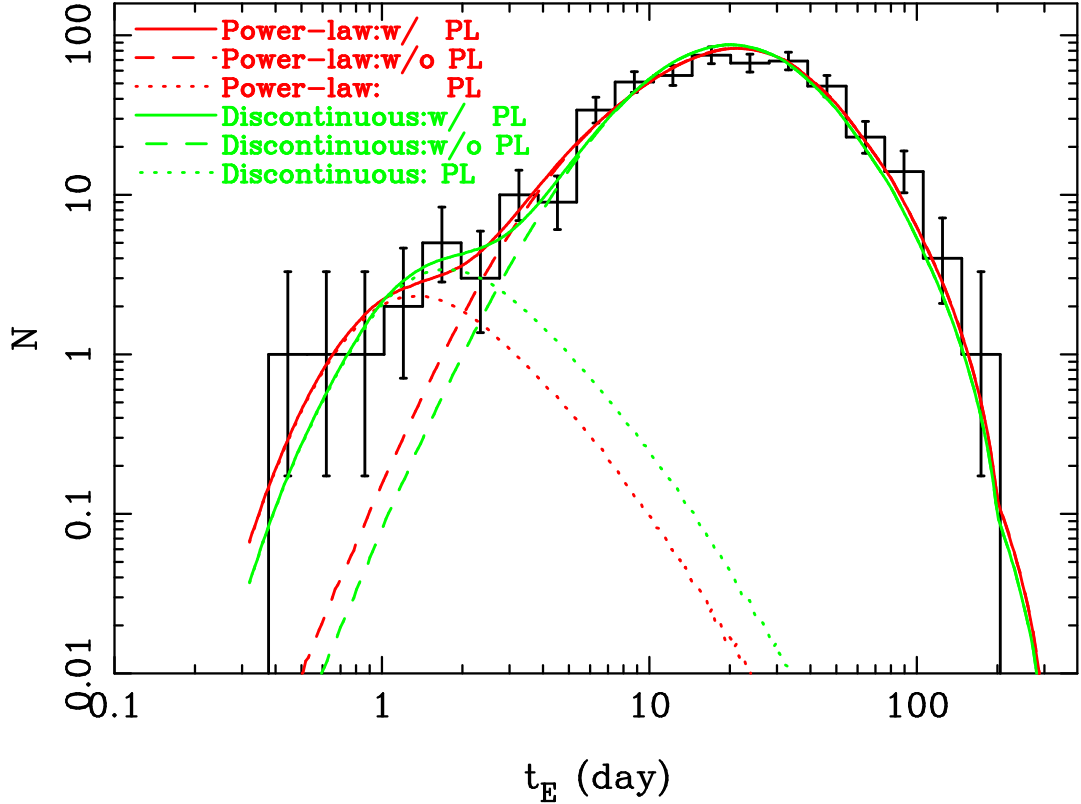}
\caption{
Observed and theoretical distributions in the timescale $t_{\rm E}$ as in
Fig. 2 of the main paper. The black histogram 
represents the observed 474 microlensing events with the error bars indicating s.e.m.. 
The green lines indicates the best-fit for the  discontinuous brown dwarf mass 
function \cite{Thies2007,Thies2008} model (3) with $R_{\rm HBL}=0.3$. The red line indicates
the model (1) Power-law mass function.
In each model, dashed lines indicate models for stellar, stellar remnant and brown
dwarf populations. Dotted lines represent the Planetary-mass population (PL).
Solid lines are the sum of these populations. Although the discontinuous
brown dwarf model is a poorer fit to the data than model (1), the fit on
the low-mass end is reasonable, and the mass function for the
planetary mass objects is 
consistent with the results for models (1) and (2), which are presented in
the main paper.
\label{fig:tE_Kroupa}}
\end{figure}

%-------------FIG.S10--------------------
\begin{figure}
%\epsscale{0.5}
\includegraphics[angle=-90,scale=0.6,keepaspectratio]{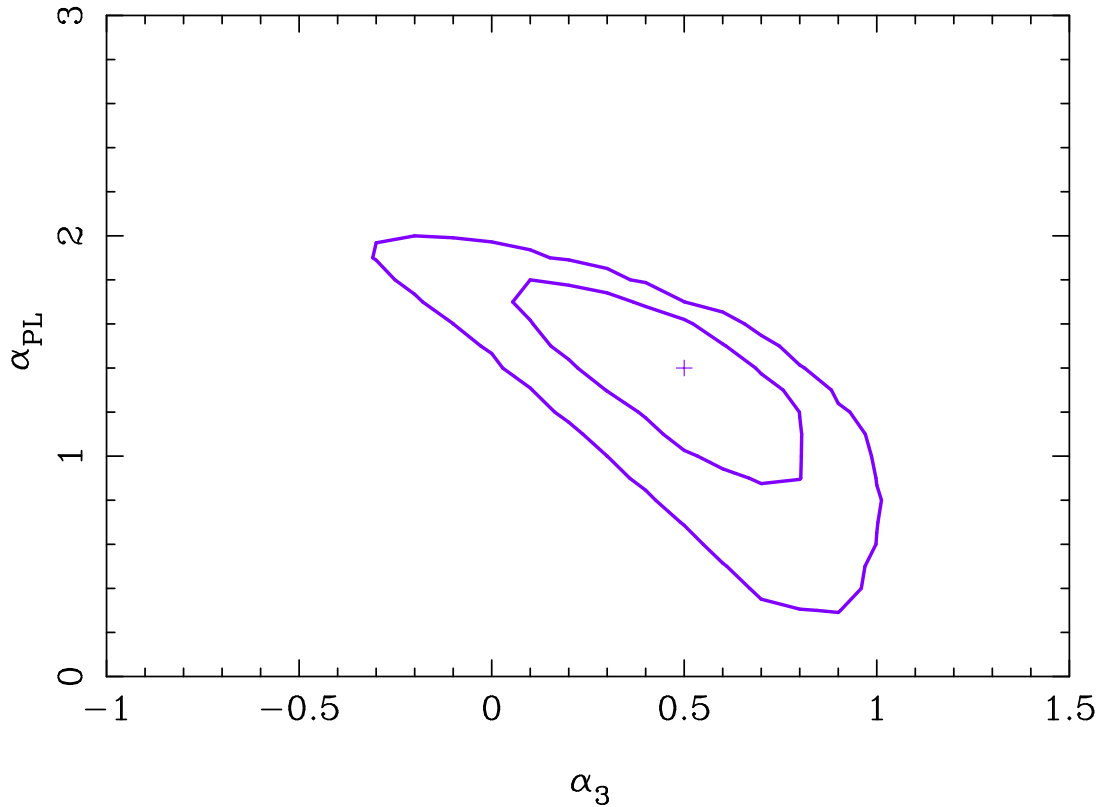}
\caption{
Likelihood contours for the Power-law indices $\alpha_3$ in the brown dwarf 
regime, and $\alpha_{\rm PL}$ in the planetary-mass regime. These brown dwarf
and planetary mas regimes span the mass ranges 
$0.01\le M/M_\sun \le 0.08$ and $10^{-5}<M/M_\sun<0.01$, respectively.
The two contours indicate the 68\% and 95\% confidence levels, and
"+" indicate the maximum likelihood point. These results
do not depend on the lower mass limit of $10^{-5}M_\sun$ as the
sensitivity to masses lower than $10^{-4} M_\sun$ is poor due to
the small detection efficiencies at low $t_{\rm E}$.
The $\alpha_3$ distribution is consistent with the $\alpha_3$ distribution for model (1)
with and without the $\delta$-function planetary mass function (see Table  \ref{tbl:MF}).
The value $\alpha_{\rm PL}$ is much steeper than $\alpha_3$,
indicating that this planetary mass objects are separate population from
brown dwarfs.
\label{fig:likelihood_PLPow}}
\end{figure}

%-------------FIG.S11-------------------- 
\begin{figure}
%\epsscale{0.5}
\includegraphics[angle=-90,scale=0.6,keepaspectratio]{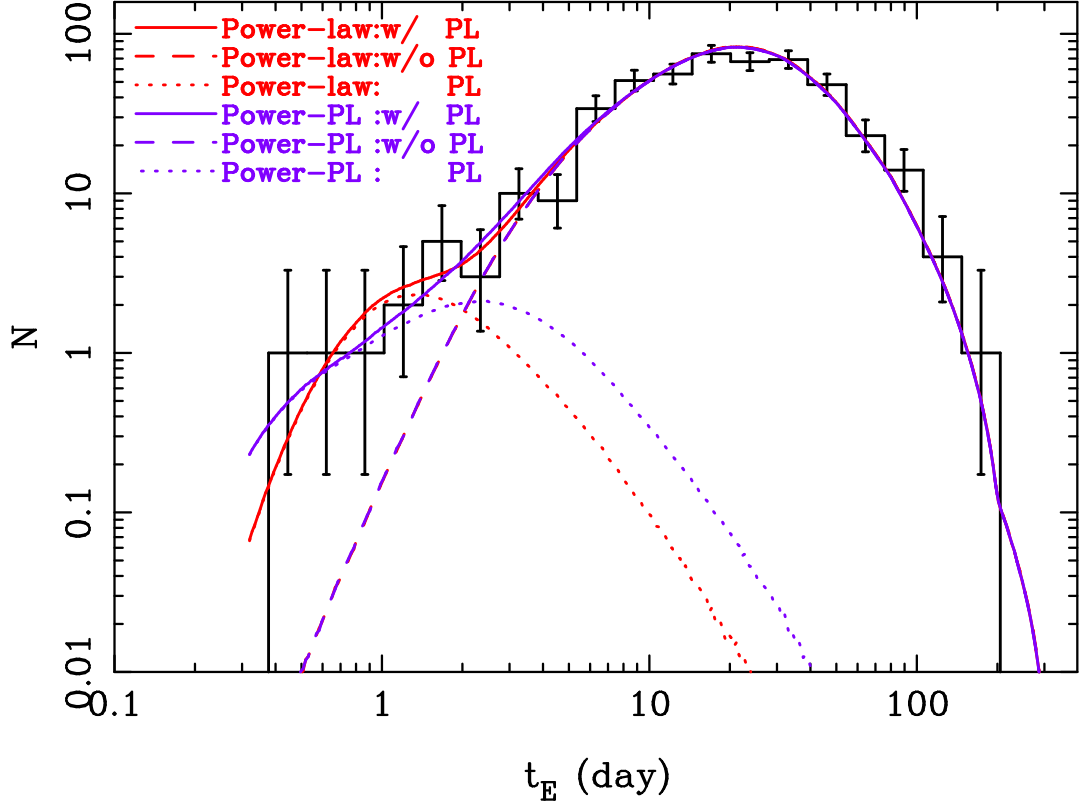}
\caption{
Observed and theoretical distributions of the timescale $t_{\rm E}$. The black histogram
represents the observed 474 microlensing events with the error bars indicating s.e.m..
The purple line indicates the best-fit model (4) with a power-law stellar and substellar 
mass function, as in model (1) , but with a continuous 
power-law mass function for in the planetary mass regime, 
$10^{-5}<M/M_\sun \le 0.08$  (Power-PL), instead of $\delta$-function mass function
of model (1).
The red line indicates the power-law mass function model (1) with a
$\delta$-function planetary mass function.
For each model, dashed lines indicate models for stellar, stellar remnant and brown
dwarf populations, and dotted lines represent the Planetary-mass population.
Solid lines are the sum of these populations. This power-law planetary
mass function model (3) has a slightly smaller likelihood value than
model (1), but because it has one fewer parameter, it is slightly favored,
formally.
\label{fig:tE_PLPow}}
\end{figure}

%-------------FIG.3--------------------
%\begin{figure}
%\epsscale{0.5}\includegraphics[angle=-90,scale=0.5,keepaspectratio]{plot_conf_contour.eps}
%\caption{
%Likelihood contours with 68\% and 95\% confidence interval on a frequency, 
%$\Phi_{\rm PL}$, and mass, $M_{\rm PL}$, of the planetary mass population 
%for the Power-law and Log-normal mass functions.  "+" indicate the maximum 
%likelihood points.
%\label{fig:likelihood}}
%\end{figure}

%-------------FIG.3--------------------
%\begin{figure}
%\includegraphics[angle=-90,scale=0.3,keepaspectratio]{plot_conf_alpha3.eps}
%\includegraphics[angle=-90,scale=0.3,keepaspectratio]{plot_conf_MPL.eps}
%\includegraphics[angle=-90,scale=0.3,keepaspectratio]{plot_conf_PhiPL.eps}
%\caption{
%Likelihood distribution with a median and 68\% and 95\% confidence interval on $\alpha_3$, 
%%%a frequency, $\Phi_{\rm PL}$, and mass, $M_{\rm PL}$
%%of the planetary mass population for the Power-law.
%\label{fig:likelihood_distribution}}
%\end{figure}

%-----------------------------------------------------
\begin{deluxetable}{lllr}
%\rotate
\tablecaption{MOA-II Galactic bulge fields\label{tbl:fields}}
\tablewidth{0pt}
\tablehead{
\colhead{Field}  & \colhead{R.A.(2000)} & \colhead{Dec.(2000)} & \colhead{$N_{\rm frame}$} \\
}
\startdata 
 gb1 & 17:47:31.41 &  -34:14:31.09 & 2,253\\
 gb2 & 17:54:01.41 &  -34:29:31.09 & 2,386\\
 gb3 & 17:54:01.41 &  -32:44:31.09 & 2,067\\
 gb4 & 17:54:01.41 &  -30:59:31.09 & 2,985\\
 gb5 & 17:54:01.41 &  -29:14:31.09 & 8,229\\
 gb6 & 17:54:01.41 &  -27:29:31.09 & 1,779\\
 gb7 & 18:00:01.41 &  -32:44:31.09 & 1,970\\
 gb8 & 18:00:01.41 &  -30:59:31.09 & 2,139\\
 gb9 & 18:00:01.41 &  -29:14:31.09 & 8,301\\
gb10 & 18:00:01.41 &  -27:29:31.09 & 1,992\\
gb11 & 18:06:01.41 &  -32:44:31.09 & 2,004\\
gb12 & 18:06:01.41 &  -30:59:31.09 & 1,790\\
gb13 & 18:06:01.41 &  -29:14:31.09 & 1,811\\
gb14 & 18:06:01.41 &  -27:29:31.09 & 1,770\\
gb15 & 18:06:01.41 &  -25:44:31.09 & 1,952\\
gb16 & 18:12:01.41 &  -29:14:31.09 & 1,756\\
gb17 & 18:12:01.41 &  -27:29:31.09 & 1,792\\
gb18 & 18:12:01.41 &  -25:44:31.09 & 1,799\\
gb19 & 18:18:01.41 &  -25:29:31.09 & 1,704\\
gb20 & 18:18:01.41 &  -23:44:31.09 & 1,679\\
gb21 & 18:18:01.41 &  -21:59:31.09 & 1,659\\
gb22 & 18:36:25.41 &  -23:53:31.09 & 1,782\\
\enddata
\tablecomments{ $N_{\rm frame}$ indicates the number of observed frames (exposures) 
during 2006-2007, which are used in this analysis. Fields gb5 and gb9 were observed 
with a 10 minute cadence, while others were observed with a 50 minute cadence. 
%The entire field gb22 and some fraction of other fields were not used in the 
%analysis because a clear Red Clump Giant population in the galactic bulge 
%are not identified.
}
\end{deluxetable}

%-----------------------------------------------------

\begin{deluxetable}{lll}
%\rotate
\tablecaption{Event Selection Criteria \label{tbl:criteria}}
\tablewidth{0pt}
\tablehead{
\colhead{level}  & \colhead{criteria} & \colhead{comments}  \\
}
\startdata  
cut0 & $N_{\rm detect}\ge3$               & Number of frames in which the object is detected. \\
\hline
cut1 &$N_{\rm data}  \ge 500$             & Number of data points\\  
&$N_{\rm out}  \ge 10$                    & Number of data points outside of the 120-day window\\  
&$\chi^2_{\rm out}/{\rm dof} \le 3$    & $\chi^2$ outside of the 120-day window\\  
%&$\sigma_{\rm max} > 10$                  & the significance of the most significant point\\  
&$N_{\rm bump} \ge 1$                     & Number of bumps in the window, where a bump\\  
&                                         & has $> 3$ consecutive points $> 3\sigma'$ above baseline\\ 
&$\chi_{3+} = \Sigma_i \left( F_i - F_{\rm base}  
\right)/\sigma'_i \ge 80$                 & Total significance of consecutive points with $> 3\sigma'$\\  
\hline
cut2 & fitting converged                 & Fits never converge if parameters are degenerate\\
& $\chi^2/{\rm dof} \le 2$            & $\chi^2$ for all data\\  
& $\chi^2_1/{\rm dof} \le 2$          & $\chi^2$ for $|t| \le t_{\rm E}$ \\  
& $\chi^2_2/{\rm dof} \le 2$          & $\chi^2$ for $|t| \le 2t_{\rm E}$ \\  
& $ 0.3 \le t_{\rm E} \le 200$ days      & Einstein radius crossing timescale\\  
& $\sigma_{t_{\rm E}}/t_{\rm E} \le 0.5$ & Error in $t_{\rm E}$\\  
& $\sigma_{t_{\rm E}} \le 12$ days       & Error in $t_{\rm E}$\\  
& $3824 \le t_0 \le 4420$ JD$'$          & Peak should be within observational period\\  
& $u_0 \le 1$                            & The minimum impact parameter\\  
& $\sigma_{u_0} \le 0.3$                 & Error in $u_0$\\  
& $I_s \le 20.0$                         & Apparent $I$-band source magnitude\\  
& $(F_{\rm s} -F_{\rm cat})/F_{\rm cat} \le 3$  & Source flux should not greatly exceed catalog flux\\  
& $\chi_{3+} \ge 70 N_{2\sigma}$ -500 & Exclude systematic residuals (depending on\\
                                                     &      & \ \ \ total significance) \\
& $\chi_{3+} \ge 45 N_{3\sigma}$ OR $N_{3\sigma} \le 2 $ & same as above\\
%\hline
\enddata
\tablecomments{
JD$'$=JD-2450000. $\sigma'_i \equiv \sigma_i \sqrt{\chi^2_{\rm out}/{\rm dof} }$.
$N_{2\sigma}$ and $N_{3\sigma}$ represent the maximum number of consecutive 
measurements which are scattered from the best fit model with an excess flux of more 
than 2-$\sigma$ and 3-$\sigma$, respectively.
}
\end{deluxetable}

\begin{deluxetable}{lllll}
%\rotate
\tablecaption{Mass Function \label{tbl:MF}}
\tablewidth{0pt}
\tablehead{
\colhead{\#}  & \colhead{Mass} & \colhead{Function} & \colhead{parameter} & \colhead{Fraction}\\
\colhead{}    & \colhead{($M_\sun$)}  & \colhead{}  & \colhead{($M$ and $\sigma$ are in $M_\sun$)}  & \colhead{$(N_*)$}\\
}
\startdata
1 & $40.0\le M$         & Gaussian           & Black hole  ($M_{\rm r}=5,   \sigma_{\rm r}=1$)         & 0.0031 \\
  & $8.00\le M\le 40.0$ & Gaussian           & Neutron star ($M_{\rm r}=1.35,\sigma_{\rm r}=0.04$)     & 0.021 \\
  & $1.00\le M\le 8.00$ & Gaussian           & White dwarf ($M_{\rm r}=0.6, \sigma_{\rm r}=0.16$)      & 0.18 \\
  & $0.70\le M\le 1.00$ & Power-law          & $\alpha_1=2.0$                              & 1.0 \\
  & $0.08\le M\le 0.70$ & Power-law          & $\alpha_2=1.3$                              &       \\
  & $0.01\le M\le 0.08$ & Power-law$^*$      & $\alpha_3=0.48_{-0.37}^{+0.29}$ w/o PL   & 0.73$_{-0.19}^{+0.22}$\\
  & $0.01\le M\le 0.08$ & Power-law$^{**}$   & $\alpha_3=0.50_{-0.60}^{+0.36}$ w/  PL   & 0.74$_{-0.27}^{+0.30}$\\
  & $M=M_{\rm PL}$      & $\delta$-function$^{**}$ & $M_{\rm PL} =1.1_{-0.6}^{+1.2} \times 10^{-3}$,$\Phi_{\rm PL}=0.49_{-0.13}^{+0.13}$  & 1.9$_{-0.8}^{+1.3}$ \\
\hline
2 & $40.0\le M$         & Gaussian       & Black hole   ($M_{\rm r}=5,   \sigma_{\rm r}=1$)       & 0.0031 \\
  & $8.00\le M\le 40.0$ & Gaussian       & Neutron star ($M_{\rm r}=1.35,\sigma_{\rm r}=0.04$)    & 0.021 \\
  & $1.00\le M\le 8.00$ & Gaussian       & White dwarf ($M_{\rm r}=0.6, \sigma_{\rm r}=0.16$)     & 0.18 \\
  & $0.08\le M\le 1.00$ & Log-normal$^*$ & $M_{\rm c} = 0.12_{-0.03}^{+0.03}$, $\sigma_{\rm c} = 0.76_{-0.16}^{+0.27}$  & 1.0\\
  & $0.01\le M\le 0.08$ & Log-normal$^*$ & $M_{\rm c} = 0.12_{-0.03}^{+0.03}$, $\sigma_{\rm c} = 0.76_{-0.16}^{+0.27}$  & 0.70$_{-0.30}^{+0.19}$\\
  & $0.00\le M\le 0.01$ & Log-normal$^*$ & $M_{\rm c} = 0.12_{-0.03}^{+0.03}$, $\sigma_{\rm c} = 0.76_{-0.16}^{+0.27}$  & 0.17$_{-0.15}^{+0.24}$\\
  & $M=M_{\rm PL}$      & $\delta$-function$^{***}$ &  $M_{\rm PL} =0.83_{-0.51}^{+0.96} \times 10^{-3},\Phi_{\rm PL}=0.46_{-0.15}^{+0.17}$  & 1.8$_{-0.8}^{+1.7}$\\
\hline
3 & $40.0\le M$          & Gaussian              & Black hole  ($M_{\rm r}=5,   \sigma_{\rm r}=1$)       & 0.00060 \\
  & $8.00\le M\le 40.0$  & Gaussian              & Neutron star ($M_{\rm r}=1.35,\sigma_{\rm r}=0.04$)     & 0.0061 \\
  & $1.00\le M\le 8.00$  & Gaussian              & White dwarf ($M_{\rm r}=0.6, \sigma_{\rm r}=0.16$)      & 0.097 \\
  & $0.50\le M\le 1.00$  & Power-law             & $\alpha_1=2.3$                              & 1.0 \\
  & $0.075\le M\le 0.50$ & Power-law             & $\alpha_2=1.3$                             &       \\
  & $0.01\le M\le 0.075$ & Power-law             & $\alpha_3=0.3$, $R_{\rm HBL}=0.3$           & 0.19\\
  & $M=M_{\rm PL}$       & $\delta$-function     & $M_{\rm PL} =1.9_{-0.9}^{+1.4} \times 10^{-3},\Phi_{\rm PL}=0.50_{-0.10}^{+0.11}$  & 1.3$_{-0.4}^{+0.7}$\\
\hline
4 & $0.08\le M  $          &                    &  same as model (1)                        & \\
  & $0.01\le M\le 0.08$    & Power-law$^{**}$   & $\alpha_3=0.49_{-0.27}^{+0.24}$ w/  PL   & 0.73$_{-0.15}^{+0.17}$\\
  & $10^{-5}\le M\le 0.01$ & Power-law$^{**}$   & $\alpha_{\rm PL}=1.3_{-0.4}^{+0.3}$ w/  PL  & 5.5$_{-4.3}^{+18.1}$ \\
\enddata
\tablecomments{
\footnotesize
\# is the model ID number:  (1) power-law+$\delta$-function, (2) log-normal+$\delta$-function, 
(3) discontinuous power-law+$\delta$-function and (4) power-law+power-law.
Gaussian: $dN/dM=\exp[(M- M_{\rm r})^2/(2\sigma_{\rm r}^2)]$. Function types are: 
Power-law: $dN/d\log M=M^{1-\alpha}$. 
Log-normal:  $dN/d\log M=\exp[(\log M-\log M_{\rm c})^2/(2\sigma_{\rm c}^2)]$. 
Fraction is the number of objects relative to the number of main sequence 
stars with $0.08\le M/M_\sun\le 1$, $N_*$ ($0.075\le M/M_\sun\le 1$ for the model 3). 
The number of stellar remnants is estimated by extending the upper main sequence 
Power-law $\alpha=2.0$ ($\alpha=2.3$ for the model 3) through this higher mass regime.
$*$: $\alpha_3$ or ($M_{\rm PL}$, $\Phi_{\rm PL}$) are fit to events with $t_{\rm E}>2$ days 
without the planetary mass $\delta$-function.
$**$: $\alpha_3$ and ($M_{\rm PL}$, $\Phi_{\rm PL}$) ($\alpha_{\rm PL}$ for the model 3) are 
fit simultaneously for the full sample.
$***$: $M_{\rm c}=0.12$ and $\sigma_{\rm c}=0.76$ are held fixed when fitting $M_{\rm PL}$ 
and $\Phi_{\rm PL}$. The fraction of planetary mass objects in model (4) is large because 
it extends down to $M=10^{-5} M_\sun$ where have no sensitivity.
}
\end{deluxetable}

\end{document}